\documentclass[10pt,aps,prd,preprintnumbers,twocolumn,nofootinbib,floatfix,a4paper,latexsym,array,enumerate,amsmath,amssymb,superscriptaddress]{revtex4-2}

\usepackage{graphicx} 
\usepackage{dcolumn}
\usepackage{xcolor}
\usepackage[colorlinks,linkcolor=blue,anchorcolor=blue,citecolor=blue,urlcolor=blue]{hyperref}
\usepackage{orcidlink}
\usepackage[mathlines]{lineno}
\usepackage{multirow}
\usepackage{threeparttable}
\usepackage{CJK}

\usepackage{newtxtext}
\usepackage{newtxmath}

\usepackage{todonotes}
\setuptodonotes{color=blue!30}

\newcommand{\be}{\begin{eqnarray}}
\newcommand{\ee}{\end{eqnarray}}
\newcommand{\bea}{\begin{eqnarray}}
\newcommand{\eea}{\end{eqnarray}}
\newcommand{\beq}{\begin{eqnarray}}
\newcommand{\eeq}{\end{eqnarray}}

\newcommand{\bef}[1][]{\begin{figure*}[#1]}
\newcommand{\eef}{\end{figure*}}

\newcommand{\bet}[1][]{\begin{table*}[#1]} 
\newcommand{\eet}{\end{table*}}

\newcommand{\Fig}[1]{Fig.~\ref{#1}}

\newcommand{\mrm}[1]{\mathrm{#1}}
\newcommand{\eV}{\mathrm{eV}}
\newcommand{\Hz}{\mathrm{Hz}}
\newcommand{\cm}{\mathrm{cm}}

\newcommand{\kpc}{\mathrm{kpc}}

\newcommand{\uni}{\mrm{U}}
\newcommand{\logu}{\mrm{log}_{10}$-$\mrm{U}}

\begin{document}

\title{Constraints on Ultralight Scalar and Dark Photon Dark Matter from PPTA-DR3 and EPTA-DR2}

\author{Xiao-Song Hu
\orcidlink{0009-0005-2784-4961} }
\affiliation{School of Physics and Astronomy, Beijing Normal University, Beijing 100875, China}
\affiliation{Department of Physics, Faculty of Arts and Sciences, Beijing Normal University, Zhuhai 519087, China}
\affiliation{State Key Laboratory of Radio Astronomy and Technology, Shanghai Astronomical Observatory, CAS, Shanghai 200030, China}

\author{Siyuan Chen
\orcidlink{0000-0002-3118-5963} }
\email{siyuan.chen@shao.ac.cn}
\affiliation{State Key Laboratory of Radio Astronomy and Technology, Shanghai Astronomical Observatory, CAS, Shanghai 200030, China}

\author{Kuo Liu
\orcidlink{0000-0002-2953-7376} }
\email{liukuo@shao.ac.cn}
\affiliation{State Key Laboratory of Radio Astronomy and Technology, Shanghai Astronomical Observatory, CAS, Shanghai 200030, China}

\author{Xingjiang Zhu
\orcidlink{0000-0001-7049-6468} }
\email{zhuxj@bnu.edu.cn}
\affiliation{Department of Physics, Faculty of Arts and Sciences, Beijing Normal University, Zhuhai 519087, China}
\affiliation{Institute for Frontier in Astronomy and Astrophysics, Beijing Normal University, Beijing 102206, China}

\author{Shi-Yi Zhao
\orcidlink{0009-0001-8885-5059} }
\affiliation{School of Physics and Astronomy, Beijing Normal University, Beijing 100875, China}
\affiliation{Department of Physics, Faculty of Arts and Sciences, Beijing Normal University, Zhuhai 519087, China}

\author{Wu Jiang
\orcidlink{0000-0001-7369-3539} }
\affiliation{State Key Laboratory of Radio Astronomy and Technology, Shanghai Astronomical Observatory, CAS, Shanghai 200030, China}


\author{John Antoniadis
\orcidlink{0000-0003-4453-3776}}
\affiliation{Institute of Astrophysics, Foundation for Research and Technology -- Hellas (FORTH), GR-70013 Heraklion, Greece}

\author{N. D. Ramesh Bhat
\orcidlink{0000-0002-8383-5059}}
\affiliation{International Centre for Radio Astronomy Research, Curtin University, Bentley, WA 6102, Australia}

\author{Amodio Carleo
\orcidlink{0000-0001-9929-2370}}
\affiliation{INAF, Osservatorio Astronomico di Cagliari, Via della Scienza 5, 09047 Selargius (CA), Italy}

\author{Shi Dai
\orcidlink{0000-0002-9618-2499}}
\affiliation{Australia Telescope National Facility, CSIRO, Space and Astronomy, PO Box 76, Epping, 1710, NSW, Australia}

\author{Valentina Di Marco
\orcidlink{0000-0003-3432-0494}}
\affiliation{School of Physics, University of Melbourne, Parkville, VIC 3010, Australia}
\affiliation{ARC Centre of Excellence for Gravitational Wave Discovery (OzGrav)} 

\author{Huanchen Hu
\orcidlink{0000-0002-3407-8071}}
\affiliation{Lohrmann Observatory, Technische Universit\"{a}t Dresden, Mommsenstra\ss e 13, 01062 Dresden, Germany}
\affiliation{Max-Planck-Institut f\"{u}r Radioastronomie, Auf dem H\"{u}gel 69, 53121 Bonn, Germany}

\author{Wenhua Ling
\orcidlink{0009-0009-9142-6608}}
\affiliation{Australia Telescope National Facility, CSIRO, Space and Astronomy, PO Box 76, Epping, 1710, NSW, Australia}

\author{Yang Liu
\orcidlink{0000-0003-0713-6640}}
\affiliation{State Key Laboratory of Radio Astronomy and Technology, Shanghai Astronomical Observatory, CAS, Shanghai 200030, China}

\author{Saurav Mishra
\orcidlink{0009-0001-5633-3512}}
\affiliation{Centre for Astrophysics and Supercomputing, Swinburne University of Technology, Hawthorn, VIC, 3122, Australia} 
\affiliation{ARC Centre of Excellence for Gravitational Wave Discovery (OzGrav)}
\affiliation{Australia Telescope National Facility, CSIRO, Space and Astronomy, PO Box 76, Epping, 1710, NSW, Australia}

\author{Christopher J Russell
\orcidlink{0000-0002-1942-7296}}
\affiliation{CSIRO Scientific Computing, Australian Technology Park, Locked Bag 9013, Alexandria, NSW 1435, Australia}

\author{Ryan M. Shannon
\orcidlink{0000-0002-7285-6348}}
\affiliation{Centre for Astrophysics and Supercomputing, Swinburne University of Technology, Hawthorn, VIC, 3122, Australia}
\affiliation{ARC Centre of Excellence for Gravitational Wave Discovery (OzGrav)}

\author{Clemente Smarra
\orcidlink{0000-0002-0817-2830}}
\affiliation{Dipartimento di Fisica e Astronomia ‘G. Galilei’, Università di Padova, Via F. Marzolo 8, 35131 Padova, Italy}
\affiliation{INFN Sezione di Padova, Via F. Marzolo 8, 35131 Padova, Italy}

\author{Jingbo Wang
\orcidlink{0000-0001-9782-1603}}
\affiliation{Institute of Optoelectronic Technology, Lishui University, Lishui 323000, China}

\author{Lin Wang
\orcidlink{0000-0003-0757-3584}}
\affiliation{State Key Laboratory of Radio Astronomy and Technology, Shanghai Astronomical Observatory, CAS, Shanghai 200030, China}

\author{Andrew Zic
\orcidlink{0000-0002-9583-2947}}
\affiliation{Australia Telescope National Facility, CSIRO, Space and Astronomy, PO Box 76, Epping, 1710, NSW, Australia}
\affiliation{ARC Centre of Excellence for Gravitational Wave Discovery (OzGrav)}

\date{\today}



\begin{abstract}
The cold dark matter model successfully describes the Universe on large scales, yet faces challenges at sub-galactic scales. Ultralight dark matter (ULDM), with particle masses around $10^{-22} \ \eV$, offers a promising solution to these small-scale issues. Pulsar Timing Arrays (PTAs), designed to detect nanohertz gravitational waves, can also provide a sensitive probe for ULDM signals. In this work, we perform a Bayesian search for ULDM using PTA data sets, focusing on two types of signals: the oscillatory gravitational potential from scalar ULDM and the fifth-force interaction mediated by dark photon dark matter (DPDM). We incorporate pulsar distances in the analysis to better model the ULDM density. No statistically significant evidence for ULDM has been found, therefore we place 95\% confidence-level upper limits on the relevant parameters. For scalar ULDM, our analysis does not exclude the scenario in which ULDM constitutes all of dark matter. The constraints from PPTA-DR3 show significant improvements over the earlier PPTA-DR2 (2018 Preview) across most of the mass range, and are consistent with the recent uncorrelated limits from other PTAs. We also present for the first time the DPDM constraints using EPTA data. The obtained bounds on the DPDM from the EPTA-DR2 and PPTA-DR3 are comparable to existing constraints.
\end{abstract}

\maketitle


\section{introduction}
\label{sec1:intro}

Dark matter is one of the foremost mysteries in modern physics. It accounts for roughly 27\% of the Universe’s total energy density \cite{2018_planck}, yet its fundamental nature remains elusive. The cold dark matter (CDM) paradigm is widely accepted, as it successfully accounts for a broad range of cosmological observations spanning significant redshift intervals \cite{1999_Bahcall}. Although CDM accurately describes the large-scale structure and evolution of the Universe, it continues to face persistent challenges on sub-galactic scales. These include the “core-cusp problem” \cite{1994_Flores,1994_Moore}, “missing-satellite problem” \cite{1999_Klypin,1999_Moore}, and the “too-big-to-fail problem” \cite{Boylan-Kolchin:2011qkt}, see Ref \cite{Bullock:2017xww} for a review. While some of these small-scale issues may be alleviated through baryonic feedback mechanisms in astrophysical modeling \cite{2015_Chan}, accurately capturing such processes remains challenging due to their complex dependence on galaxy formation physics. Alongside these astrophysical approaches, several particle physics frameworks offer alternative pathways to address the observed discrepancies. These include warm \cite{2001_wdm_Bode}, self-interacting \cite{2018_sfdm_Tulin}, and fuzzy dark matter \cite{2000_fdm_hu}.

Among these candidates, fuzzy dark matter has emerged as a particularly compelling scenario \cite{2000_fdm_hu,2015_Weinberg,2015_MarshDavid,2017_Hui,2021_hui}. It is a type of ultralight dark matter (ULDM) with particle masses $\sim 10^{-22} \ \eV$, its de Broglie wavelength (coherence length) extends to sub-galactic scales ($\sim 0.5 \ \kpc$). ULDM produces diverse observational effects. Its properties are constrained by various astrophysical observations such as cosmic microwave background, Lyman-$\alpha$ forest and 21-$\cm$ cosmology etc \cite{2021_Ferreira}. However, the strength of these constraints depends heavily on the modeling of small-scale structure \cite{2014_Schive,2014_Schive_3dsimulations,2018_zhangjiajun}. Therefore, an independent approach less affected by these uncertainties is highly valuable. Pulsar timing array (PTA) experiments serve as such an alternative. These experiments monitor highly stable millisecond pulsars to obtain precise measurements of their pulse times-of-arrival (TOAs) \cite{1983_Hellings}. PTA is primarily designed to detect nanohertz gravitational waves, and also offers a promising avenue to search for ULDM signals.

ULDM imprints detectable signals in PTA observations via two principal mechanisms. The first mechanism is gravitational effects \cite{2014_Andrei}. ULDM oscillations induce fluctuations in the spacetime metric. These metric variations alter the gravitational potential and generate measurable changes in the TOAs. The second mechanism involves interactions with the Standard Model \cite{Smarra_2024,2018_PPTA_darkphoton_xuexiao,2022_kaplan_clock}. ULDM couples to the matter on Earth and on the pulsars or to a reference clock. This coupling produces additional modifications in the TOAs through a distinct mechanism.

Several previous studies have employed PTAs to search for ULDM. They typically assumed that the dark matter energy density is uniform between Earth and the pulsars \cite{2014_Porayko,2018_Porayko_PPTADM,2018_PPTA_darkphoton_xuexiao,2020_katoryo,2020_NomuraKimihiro_vector,2022_PPTA_vector_wuyumei,2022_kaplan_clock,2023_EPTA_uldm,2023_nanograv_newphysics,2023_xiaziqing_gammapTA,2024_luuhoangnhan}. In this approach, the local density measured at Earth sets a common amplitude for the timing residual. Consequently, the pulsar term may be either under- or overestimated. To mitigate the impact by this issue, here we keep the common amplitude, however, treat the pulsar distances as free parameters. This allows us to compute the dark matter density ratio between Earth and the pulsars, which in turn modifies the common amplitude for the signal. Marginalizing over pulsar distance uncertainties helps to avoid potential inaccuracies in modeling the signal at the location of the pulsar.

In this work, we search for ULDM by targeting two distinct classes of signals: the oscillatory gravitational potential induced by scalar ULDM, and the fifth-force coupling arising from dark photon dark matter. We advance the methodology of \cite{2018_Porayko_PPTADM} by developing a simplified framework that incorporates pulsar-Earth term coherence as an effective parameter and employs the dark matter density ratio to modify the common signal amplitude. We analyze the PPTA-DR3 and EPTA-DR2 to search for both types of ULDM signals. The article is organized as follows. In Sec. \ref{sec2}, we introduce the theoretical framework of ULDM-induced timing residuals. Sec. \ref{sec3:analysis} details the data, noise analysis and Bayesian methodology. Our results, the corresponding discussion and conclusions are presented in Sec. \ref{sec4:results}.


\section{Ultralight dark matter}
\label{sec2}

ULDM is characterized by an extremely high occupation number and non-relativistic velocity ($v_{0} \sim 10^{-3}\ c$), which allows for its treatment as a classical wave field. 
A typical wavemode of ULDM can be written as  \cite{2014_Andrei,2018_Foster,2021_Foster},

\beq
\phi(x,t) = A  \hat{\phi}(x) \sin( \omega t - \mathbf{k} \cdot x+ \alpha(x))
\eeq

where $A$ is the amplitude determined by the dark matter density and mass. The factor $\hat{\phi}(x)$ captures the spatial pattern of stochastic fluctuations arising from the wave-like nature of ULDM \cite{2019_Centers}. The angular frequency (particle's energy) is $\omega = m + {|\mathbf{k}|^2}/{(2m)}$. For a virialized halo with characteristic velocity $v_0$, the characteristic momentum is $|\mathbf{k}| = m \mathbf{v_{0}}$ and the corresponding characteristic frequency is $\omega \approx m(1+v_0^2/2)$. Here, $m$ is the ULDM particle mass and $x$ is the position and $\alpha $ is a random phase. Two distinct regimes are relevant: the fast mode with $\omega  \approx m$, corresponding to coherent oscillations near the Compton frequency, and the slow modes constrain all the frquency $\omega \lesssim mv_{0}^2/2$, characterizing low-frequency stochastic variations. PTA are sensitive to metric oscillations in the nHz band, corresponding to ULDM mass range of approximately $ 10^{-24} \ \eV-10^{-22} \ \eV$ for the fast model and $10^{-18} \ \eV-10^{-16} \ \eV$ for the slow model. In this analysis, we focus on the fast mode, to which PTA exhibit particularly compelling sensitivity \footnote{The slow mode originates from higher-order terms $\omega \lesssim mv_0^2/2$, which were neglected in \cite{2014_Andrei} for the dark matter mass around $10^{-23} \ \eV$ due to their large length scale (Mpc). In the mass range $10^{-18} - 10^{-16} \ \eV$, the slow model produce characteristic low-frequency fluctuations and generate additional pulsar cross-correlations \cite{2023_kim,2024_Kim_pta}. 
However, the amplitudes of both fast and slow modes scale as $A \propto \sqrt{\rho}/m$, where $\rho$ is the dark matter energy density. Under the same strain sensitivity, the resulting upper limits on the dark matter density weaken for heavier ULDM particles due to the inverse proportionality between the signal amplitude and the particle mass. 
Consequently, the slow modes constraints are systematically weaker than their fast mode counterparts.}. 

The coherence properties of the ULDM field are characterized by the coherence time $\tau_c$ and coherence length $l_c$. The coherence time scales as $\tau_c \sim 2\pi/mv_{0}^2$. With particle masses $m\lesssim 10^{-22} \ \eV$, the coherence time $\tau_{c} \gtrsim 10^6 $ yr, far exceeding typical observational baselines. ULDM exhibits coherent oscillations on spatial scales smaller than the coherence length $l_c$, the coherence length $l_c$ is,

\beq
l_c = \frac{2 \pi }{m v_{0}} \approx 0.4 \ \kpc \left ( \frac{m}{10^{-22} \ \eV} \right)^{-1} 
\eeq

In this work, we consider two simplified regimes based on the relative scale of the dark matter coherence length: the uncorrelated case and the correlated case. The strength of the ULDM field at Earth and the pulsar are denoted by $\hat{\phi}_e$ and $\hat{\phi}_p$ respectively.

In the uncorrelated case, the coherence length is smaller than typical Earth-pulsar and pulsar-pulsar distances ($l_c \ll |x_{(e,p)}|$); $\hat{\phi}_e$ and $\hat{\phi}_p$ are treated as independent parameters, following the approach adopted in \cite{2023_EPTA_uldm,2023_nanograv_newphysics}. Independent priors on $\hat{\phi}_e$ and $\hat{\phi}_p$ can lead to numerically stable convergence of the posteriors, but the resulting estimates remain highly sensitive to the choice of the priors.

In the correlated case, the coherence length exceeds typical Earth-pulsar and pulsar-pulsar distances ($l_c \gg |x_{(e,p)}|$), leading to $\hat{\phi}_e = \hat{\phi}_p = \hat{\phi}$. The EPTA-DR2 analysis \cite{2023_EPTA_uldm} further distinguishes two sub‑cases in this regime based on the scale of the Milky Way rotation curve: “pulsar-correlated”, where $l_c$ remains smaller than that scale and $\hat{\phi}$ is treated independently and “correlated” where $l_c$ exceeds that scale and $\hat{\phi}$ is absorbed into $A = \hat{\phi} \times A$. Rather than treating $\hat{\phi}$ as an independent parameter, we absorb it into $A$. This choice is motivated by two considerations: First, for signal detection, the uncorrelated case already includes $\hat{\phi}_e = \hat{\phi}_p$ as a special case. Second, treating $\hat{\phi}$ independently introduces a strict degeneracy, because the product $\hat{\phi} \times A$ is invariant under a rescaling of one parameter and an inverse rescaling of the other. Given the considerations above, we do not further subdivide the correlated regime according to the scale of the Milky Way rotation curve and simply retain the two‑regime description. In contrast, although the NANOGrav analysis \cite{2023_nanograv_newphysics} adopts the same physical coherence condition, it retains $\hat{\phi}$ explicitly rather than absorbing it into $A$, while still labeling the scenario as “correlated”.

The ULDM wave field is stochastic, which arises from the interference among field modes oscillating at slightly different frequencies due to the small velocity dispersion. To correctly model its stochastic nature, we represent the ULDM using an exponential prior \cite{2018_Foster,2019_Centers}. 

The amplitude of ULDM signal depends on the local dark matter density. Previous studies commonly assumed a uniform density between Earth and all pulsars, actual densities likely vary from pulsar to pulsar. Such variations would modulate the oscillation amplitude, potentially leading to an under- or overestimation of the pulsar term in the timing residual. To incorporate this effect, we introduce a pulsar-to-Earth density ratio $\delta {\rho_\mrm{DM}}$, it can be written as, 

\beq
\delta \rho_{\mrm{DM}} = \frac{\rho(x_p)}{\rho(x_e)}
\eeq

where $\rho (x)$ is the dark matter energy density at the corresponding location. ULDM forms solitonic core structures in the inner regions of dark matter halos due to quantum pressure. These halos exhibit a dense, stable solitonic core in the central region, while the outer structure can be described by Navarro–Frenk–White (NFW) profile \cite{1996_profile,2024_Liao}. It should be noted that a unified numerical study covering the full range of possible particle masses is still lacking in recent research \cite{2025_Schive}. In the absence of such a complete model, 
we adopt the soliton + NFW profile as a simplified and observationally consistent description of the overall halo density \cite{2023_liutao},

\beq
\rho(r)= \kappa \times
\begin{cases} 
\frac{0.019(\frac{m_{a}}{m_{22}})^{-2}(\frac{r_{c}}{\mathrm{kpc}})^{-4}}{[1+9.1\times10^{-2}(\frac{r}{r_{c}})^{2}]^{8}} 
& \mathrm{for~}r<l_{c} \\ 
\frac{\rho_{s}}{r/r_s(1+r/r_s)^{2}}, & \mathrm{for~}r>l_{c}  
\end{cases}
\eeq
where $\kappa = \Omega_{\mathrm{ULDM}} / \Omega_{\mathrm{DM}}$ defines the fractional relic abundance for ULDM. 
Under this dark matter profile, for the correlated case, $\delta \rho$ depends on the ULDM coherence length. As it becomes sufficiently large to encompass the Earth-pulsar separations, we can approximate $\delta \rho \simeq 1$. For the uncorrelated case, the Earth and pulsars are instead described by the NFW profile. Here, the $\rho_s = 0.471 \ \mrm{GeV/cm^3}$ is the characteristic density, the $r_s = 16.1 \ \kpc$ is the characteristic radius of the dark matter halo \cite{2013_Nesti}. At the Earth's position in the Galaxy, where the distance to Galactic center $x_e = 8.3 \ \kpc$, the local dark matter energy density is $\rho_0 \approx 0.4 \ \mrm{GeV/cm^3}$ \cite{2012_bovy,2014_Read,2018_Sivertsson,2020_desalas}. 

The dark matter density ratio between Earth and a pulsar is determined by their respective positions within the NFW profile. However, pulsar distances from Earth remain subject to significant uncertainties \cite{pptadr3_data}. To account for these uncertainties, we incorporate pulsar distances as free parameters in our Bayesian analysis, following a strategy commonly employed in continuous gravitational wave searches \cite{12.5cw_nanograv,15cw_nanograv,Zhao_2025}. We describe the pulsar distance prior in Appendix \ref{pulsar_distance}, with specific distance values and prior distributions listed in Appendix Table \ref{pulsar_distance_prior}. The derived dark matter density ratio $\delta\rho_{\mathrm{DM}}$, with uncertainties propagated from the distance, ranges from approximately 0.8 to 1.6 as shown in \Fig{S1_delta_rho_DM}. This spread indicates spatial variations in the local dark matter density between Earth and the pulsars.

\bef[t]
\centering
\includegraphics[width=0.9\textwidth]{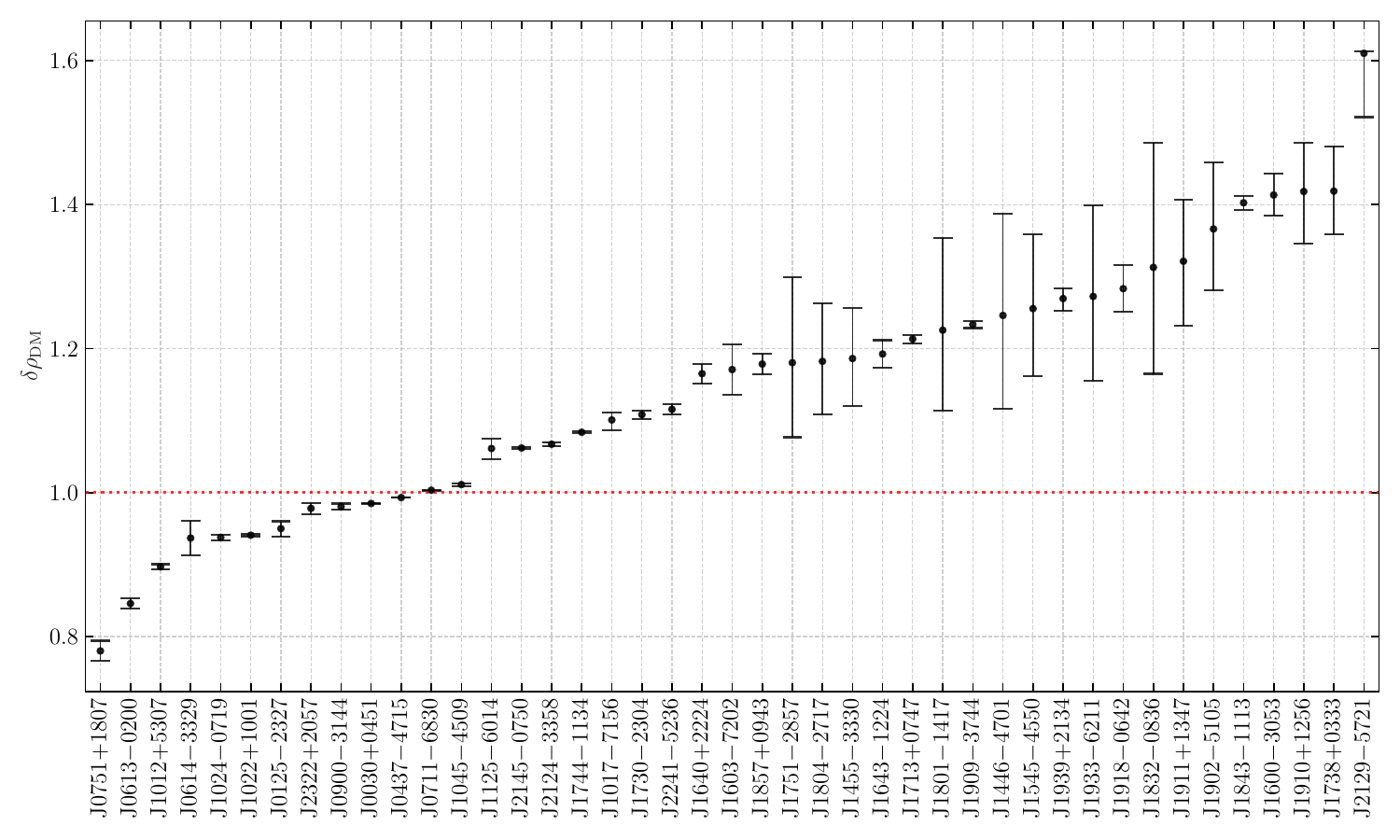}
\caption{The ratio of the dark matter density at each pulsar to that at Earth under the NFW profile for the uncorrelated case is presented, with measurement uncertainties indicated by error bars. The red dashed line corresponds to equal density ($\delta\rho_{\mrm{DM}} = 1$).}
\label{S1_delta_rho_DM}
\eef


\subsection{Scalar ULDM: Gravitational Signal}
ULDM induces periodic oscillations in the spacetime metric between Earth and the pulsars, modifying the gravitational potential and thereby producing characteristic variations in the measured TOAs \cite{2014_Andrei}. In this work, we focus exclusively on the gravitational signal of scalar ULDM \footnote{
Signals from vector and tensor ULDM involve additional degrees of freedom and richer phenomenology \cite{2020_NomuraKimihiro_vector,2023_Omiya,2018_Marzola,2024_cai}, making their analysis significantly more complex. 
}. The resulting timing residuals for scalar ULDM can be expressed as \cite{2014_Andrei},

\beq
\begin{aligned}
\Delta t= \frac{\Psi}{2 m} & [ \hat{\phi_{e}^2} \sin(2 m t + \alpha_e) - \\ & \delta \rho_{\mrm{DM}} \hat{\phi_{p}^2} \sin(2 m t + \alpha_p -2md_p/c)]
\end{aligned}
\label{time_residual_01}
\eeq 

$\Psi$ represents the amplitude of the oscillation of the gravitational potential. The parameters $\hat{\phi_{e}}=\hat{\phi}(x_e)$ and $\hat{\phi_{p}}=\hat{\phi}(x_p)$ are random variables characterizing the superposition of the ULDM stochastic field at the Earth and pulsar locations, respectively. For scalar ULDM with particle mass $m$, the oscillation frequency is given by $ f = 2m/2\pi \approx 4.84 \times 10^{-8} (m/10^{-22} \ \eV) \ \Hz$. We define the phase at Earth’s location as $\alpha_e = \alpha(x_e)$ and the phase at the pulsar as $\alpha_p = \alpha(x_p)$. In the correlated case, the coherence length is large than typical Earth–pulsar and pulsar–pulsar distances, leading $\alpha_e = \alpha_p$. In the uncorrelated case, $\alpha_e$ and $\alpha_p$ are independent, and the distance uncertainty phase $2 m d_p / c$ is absorbed into $\alpha_p$. The timing residual incorporates the same amplitude term through the dark matter density ratio $\delta\rho_{\mrm{DM}}$, with the gravitational potential amplitude expressed as,

\beq
\begin{aligned}
\Psi = \frac{\pi G \rho}{m^2} \approx 6.48 \times 10^{-18} \left(\frac{m}{10^{-22} \ \eV} \right)^{-2} \left(\frac{\rho}{\rho_0}\right)
\end{aligned}
\eeq

Through the amplitude term $\Psi$, the sensitivity to ULDM increases as the local dark matter density becomes larger. These findings agree with a previous study that investigated the amplitude as a function of constant distance, demonstrating that closer proximity to the Galactic center, where the dark matter density is higher, enhances the sensitivity of dark matter detection efforts \cite{2018_Porayko_PPTADM}. 

\subsection{Dark photon: Fifth-Force Signal}

Dark photon dark matter (DPDM) is a theoretical candidate for dark matter proposed by string theory compactifications \cite{2008_Burgess,2009_Goodsell,2011_Cicoli} and serves as the gauge boson of a hidden $U(1)$ interaction. When this symmetry corresponds to baryon number $U(1)_{B}$ or baryon-minus-lepton number $U(1)_{B-L}$, standard model particles including protons, neutrons, and electrons acquire effective charges that generate a fifth-force between ordinary matter objects. The DPDM field couples to the matter of pulsars and the Earth through a fifth-force interaction, generating accelerations that Doppler-shift the radio pulse arrival times and consequently modify the measured TOAs. In this analysis, we consider only the fifth-force coupling effects of DPDM, without including the gravitational perturbations in our model. This simplified approach is particularly relevant for vector fields, where the full signal modeling presents additional complexities \cite{2020_NomuraKimihiro_vector,2023_Omiya}. The timing residuals caused by fifth-force is given by\cite{2018_Pierce,2018_PPTA_darkphoton_xuexiao},

\beq
\begin{aligned}
\Delta t_{\mrm{DPDM}}  = \frac{\epsilon e}{m} A_{0} & [ \hat{\phi_{e}} \frac{q_{e}}{m_{e}}  \cos(m t+\alpha_{e})- \\ & \sqrt{\delta{\rho_{\mrm{DM}}}} \hat{\phi_{p}} \frac{q_{p}}{m_{p}} \cos(m t+\alpha_{p})]\cdot \boldsymbol{n}
\end{aligned}
\eeq

where $\epsilon$ is coupling strength of the gauge interaction and $e$ is the electromagnetic coupling constant. The amplitude $A_0$ corresponds to the gauge potential of the DPDM background. As a vector field, DPDM comprises three spatial components, with its average magnitude determined by the local dark matter density through the relation $|A_0|^2 = 2\rho_0/(3m^2)$. The periodic variation of the timing residuals from the DPDM fifth force are at frequency $f = m/2\pi \approx 2.42 \times 10^{-8} \ \Hz$. The parameters $q_{e,p}$ and $m_{e,p}$ represent the effective charge (under the $U(1)_{B}$ and $U(1)_{B-L}$) and the mass of the Earth and pulsar, respectively. For the $U(1)_{B}$ interaction, the charge-to-mass ratio $q/m$ is approximately $1 \ \mrm{GeV}^{-1}$ for both the Earth and pulsars. In the case of the $U(1)_{B-L}$ interaction, $q/m \approx 1 \ \mrm{GeV}^{-1}$ for pulsars and $0.5 \ \mrm{GeV}^{-1}$ for the Earth \cite{2018_PPTA_darkphoton_xuexiao}. $\boldsymbol{n}$ is the normalized position vector pointing from the Earth to the pulsar.

\section{data analysis}
\label{sec3:analysis}

\subsection{The datasets}

The Parkes Pulsar Timing Array Third Data Release (PPTA-DR3) \cite{pptadr3_data} consists of high-precision timing observations of 32 millisecond pulsars, conducted with the 64-m Parkes “Murriyang” radio telescope in Australia. Compared to the previous PPTA-DR2 \cite{pptadr2_data}, this release extends the timing baseline by approximately three years and includes observations collected with the new ultrawide-band low-frequency receiver \cite{ppta_uwl}, which provides a wider bandwidth than earlier instrumental configurations \cite{pptadr2_timing}. From the full sample of 32 pulsars in PPTA-DR3, two are excluded in the present analysis: PSR J1824$-$2452A, due to its strong red noise characteristics \cite{2010_Shannon}, and PSR J1741$+$1351, owing to its limited number of observations. It is noted that for PSR J0437$-$4715, approximately 3.4 years of TOAs data from the beginning of the dataset have been excised.

The European Pulsar Timing Array Second Data Release (EPTA-DR2) \cite{eptadr2_timing} includes timing data of 25 millisecond pulsars, obtained through a combination of six radio telescopes across Europe: the Effelsberg Radio Telescope (Germany), the \textit{Lovell} Telescope and Mark II Telescope (United Kingdom), the Nançay Radio Observatory (France), the Sardinia Radio Telescope (Italy), and the Westerbork Synthesis Radio Telescope (Netherlands). These telescopes (excluding the Mark II) jointly operate as the Large European Array for Pulsars, delivering an effective diameter of up to 194 m \cite{2016_Bassa}. Compared to EPTA-DR1 \cite{eptadr1_timing}, EPTA-DR2 incorporates data from a new generation of digital backends which facilitates coherent dedispersion techniques and significantly wider observing bandwidth, yielding enhanced sensitivity in the pulsar timing data. The EPTA-DR2 dataset encompasses the full data collection ({\tt{DR2full}}) spanning 24.7 years of observations, which includes a specialized subset ({\tt{DR2new}}) comprising the most recent 10.3 years of data obtained with the new backends.

\subsection{Noise analysis}

The TOAs of a pulsar can simply be written as,

\beq
\mrm{TOAs} \sim t_{\mrm{TM}} + \Delta t_{\mrm{Noise}} + \Delta t_{\mrm{GWB}} + \Delta t_{\mrm{Signal}}
\eeq

where $t_{\mrm{TM}}$ represents the timing model contribution, for which we adopt the DE440 solar system ephemeris \cite{2021_Park}. The noise term $\Delta t_{\mrm{Noise}}$ includes both white (time-uncorrelated) and red (time-correlated) noise components, along with other pulsar-specific noise features, like dispersion measurement noise. The gravitational-wave background (GWB) term $ \Delta t_{\mrm{GWB}}$ refers to the contribution from a noise process that shares a common red power spectrum across all pulsars. The signal term $\Delta t_{\mrm{Signal}}$ corresponds to timing residuals induced by the ULDM. We search for ULDM signatures in PTA by modeling the ULDM contribution as a deterministic signal in the pulsar's TOAs. We use different parameter priors for ULDM signal searches and upper limits: $\log_{10}$-uniform prior ($\logu$) for signal search and linear-uniform prior ($\uni$) for upper limits. 
The use of a separate prior for setting upper limits has been discussed in Appendix \cite{15cw_nanograv}.  In deriving the upper limits, the dark matter mass is assigned a delta-function prior, with its value fixed at the center of each mass bin. The mass bins are defined as approximately 40 logarithmically spaced intervals over the considered dark matter mass range.

For the pulsar noise $\Delta t _{\mrm{Noise}}$, the white noise component in pulsar timing residuals originates from instrumental effects and pulse phase jitter. This noise is modeled through three parameters: EFAC, which multiplicatively scales the template-fitting uncertainties; EQUAD, additional white noise in quadrature; and ECORR, accounting for noise correlated within individual observing sessions. The red noise comprises several components, such as achromatic spin noise and dispersion measurement variations. In this work, we model both components as power-law processes. Specialized noise components, including chromatic exponential dips and other specific features, are incorporated in our analysis. For detailed descriptions of the individual, customised pulsar noise models in each data set, we refer to the comprehensive analyses presented for PPTA‑DR3 \cite{pptadr3_noise} and EPTA‑DR2 \cite{2023_eptadr2_noise}. In this work, we adopt the same customised noise models as those implemented in the GWB searches by the respective collaborations \cite{pptadr3_gwb,2023_epta_gwb}.

Recent studies by major PTA collaborations have reported evidence for GWB \cite{2023_ppta_gwb,2023_nanograv_gwb,2023_epta_gwb,2023cpta,2025_mpta_gwb}. In this analysis, we remain agnostic as to its physical origin and model this common process as an uncorrelated red noise component with a power-law spectrum. We treat the spectral parameters as free parameters, adopting prior ranges of $-18 \leq \log_{10} \mrm{A} \leq -11$ for the amplitude and $0 \leq \gamma \leq 7$ for the spectral index. 

\subsection{Bayesian methodology}

The likelihood is a function of the number of timing-residual observations, the time-domain Gaussian likelihood can be written as \cite{2016_Arzoumanian,2017_taylor},

\beq
\mathcal{L}(\delta t|\theta)=\frac{\exp\left(-\frac{1}{2}(\delta t-\mu)^T\mathbf{C}^{-1}(\delta t-\mu)\right)}{\sqrt{\det(2\pi \mathbf{C})}}
\eeq

where $\delta t$ represents the observed timing residuals and $\mu$ denotes the dark matter signal. The covariance matrix $\mathbf{C}$ of the total noise is constructed as $\mathbf{C} = \mathbf{N} + \mathbf{TBT}^{T}$, where $\mathbf{N}$ represents the white noise covariance matrix, and $\mathbf{T}$ denotes the combined basis matrix incorporating both the timing model $\mathbf{M}$ and the red noise processes $\mathbf{F}$. The matrix $\mathbf{B}$ is the prior covariance matrix containing the hyperparameters of the noise models. For more details on the likelihood function, see \cite{2021_Stephen_review}.

\bet[t]
	\centering
	\footnotesize
	\caption{Summary of the main parameters and prior distributions used in the Bayesian analysis. “U” and “$\logu$” denote uniform and log$_{10}$‑uniform priors, respectively. Pulsar noise models were customised for each data set: the PPTA-DR3 model following \cite{pptadr3_noise}, and the EPTA-DR2 model following \cite{2023_eptadr2_noise}.}
	\label{prior}
	\begin{tabular}{l c c c c}
		\hline
        & \textbf{Parameter} & \textbf{Description} & \textbf{Prior} & \textbf{Comments} \\
        \hline
		\multirow{2}*{Red noise}
		& log$_{10} \mrm{A_{red}}$ & spectrum amplitude &$\logu[-20, -11]$ & one per pulsar \\
		& $\gamma_{\mrm{red}} $ & spectrum index  &$\uni[0,7]$ & one per pulsar \\
		\hline
		\multirow{7}*{Scalar ULDM}
		& $\Psi$ & amplitude &$\logu[-17, -11]$ (search)  & one per PTA\, \\
        &        &           & $\uni[10^{-17}, 10^{-11}]$ (limit)    & \\
        & $m\,[\mrm{eV}]$ & ULDM mass &$\logu[-24,-21.7]$ (search) & one per PTA\, \\
        &        &           &  delta function (limit)    & fixed at bin center \\
		& $\hat{\phi_{e}^2}$ & Earth factor & $e^{-x}$ & one per PTA\, \\
		& $\hat{\phi_{p}^2}$  & Pulsar factor & $e^{-x}$ & one per pulsar\, \\
		& $\alpha_e$ & Earth signal phase &$\uni[0,2\pi]$ & one per PTA\, \\
		& $\alpha_p$ & Pulsar signal phase &$\uni[0,2\pi]$ & one per pulsar\, \\
        & $d_p$ & Pulsar distance & PX/DM & one per pulsar\, \\
        \hline
        \multirow{7}*{DPDM}
        & $\epsilon$ & coupling factor & $\logu[-27,-21]$ (search) & one per PTA\, \\
        &        &           & $\uni[10^{-27}, 10^{-21}]$ (limit)    & \\
        & $m\, \mrm{[eV]}$ & ULDM mass & $\logu[-24,-21]$ (search) & one per PTA\, \\ 
        &        &           &  delta function (limit)    & fixed at bin center \\
        & $\hat{\phi^{i}_{e}}$ & Earth factor & $e^{-x}$ & three per PTA\, \\
	    & $\hat{\phi_{p}}$  & Pulsar factor & $e^{-x}$ & one per pulsar\, \\
        & $\alpha^{i}_e$ & Earth signal phase &$\uni[0,2\pi]$ & three per PTA\, \\
	    & $\alpha_p$ & Pulsar signal phase &$\uni[0,2\pi]$ & one per pulsar\, \\
        & $d_p$ & Pulsar distance & PX/DM & one per pulsar\, \\
		\hline
        \multirow{2}*{GWB}
		& log$_{10} \mrm{A_{gwb}}$ & spectrum amplitude &$\logu[-18, -11]$ & one per PTA \\
		& $\gamma_{\mrm{gwb}}$ & spectrum index  &$\uni[0,7]$ & one per PTA \\
		\hline
	\end{tabular}
\eet

We define the model parameters and specify their prior ranges using the {\tt{ENTERPRISE}} \cite{enterprise} and {\tt{ENTERPRISE\_EXTENSION}} \cite{enterprise_extensions} packages for the Bayesian inference. The main noise and signal parameters with their corresponding prior distributions are summarized in Table \ref{prior}. To improve computational efficiency, white noise parameters are fixed to their maximum-likelihood values obtained from single pulsar analyses. We evaluate the likelihood function under these priors and sample the posterior distributions using the parallel-tempered Markov Chain Monte Carlo algorithm implemented in {\tt{PTMCMCSAMPLER}} \cite{justin_ellis_2017_1037579}. For the ULDM signal detection search, we employ {\tt{Product Space Sampling}} within a hypermodel framework to compute Bayes factors (BF) \cite{2020_Taylor}, evaluating the evidence ratio between ULDM-containing models and the null signal hypothesis. Following this approach, a value of $\mrm{ln \ BF} \gtrsim 3$ is interpreted as positive evidence for a ULDM signal. When no statistically significant ULDM signal is detected, we compute 95\% confidence-level upper limits on the relevant parameters using standard PTMCMC sampling.


\section{results and discussion}
\label{sec4:results}


We performed Bayesian analyses on the PPTA-DR3 and EPTA-DR2 data sets, searching for signals from scalar ULDM and DPDM. Based on the methodology outlined in \cite{2018_Porayko_PPTADM}, our work implements and extends their framework for future searches. First, we incorporate dark matter coherence by considering both correlated and uncorrelated regimes. Second, we model the pulsar term by incorporating pulsar distances as free parameters and using the dark matter profile to compute dark matter density, setting a common signal amplitude. No statistically significant signals were identified in our search. While a modest excess around $10^{-7} \ \Hz$ was found in EPTA-DR2 ({\tt{DR2full}}), no such signal was found in EPTA-DR2 ({\tt{DR2new}}). This likely originates from unmodeled high-frequency noise in some pulsars, thus rendering the signal insignificant for further physical interpretation. We therefore focus on reporting the 95\% confidence-level upper limits on the parameters of scalar ULDM and DPDM. Detailed results for the EPTA-DR2 ({\tt{DR2full, DR2new}}) dataset are presented in Appendix \ref{eptadr2_compare}.

\subsection{Upper limits}

\bef[h]
\centering
\includegraphics[width=0.45\textwidth]{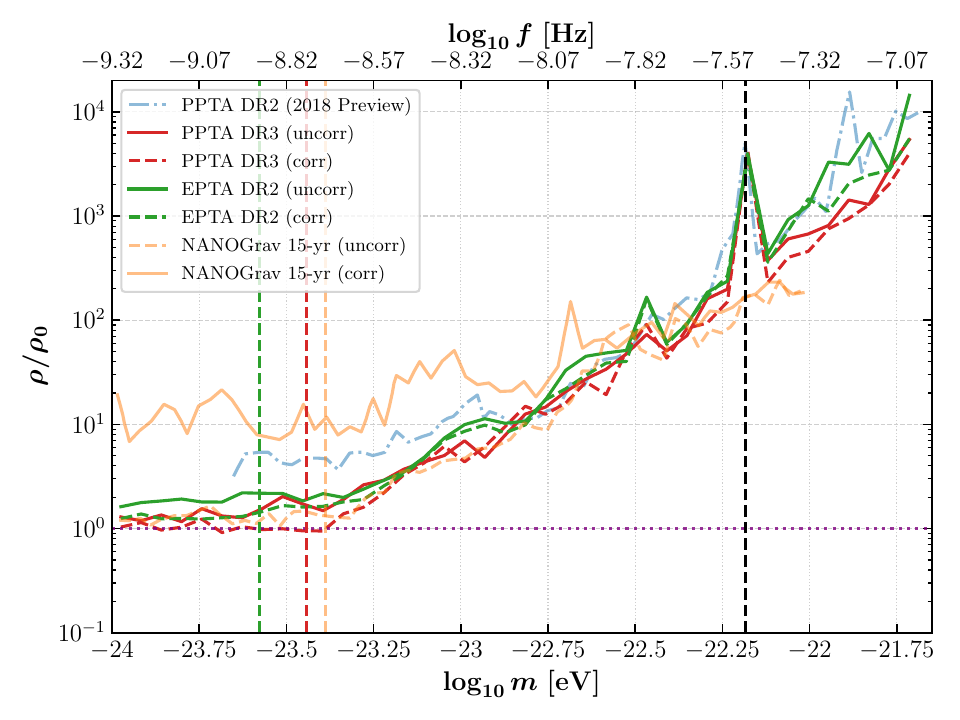}
\includegraphics[width=0.45\textwidth]{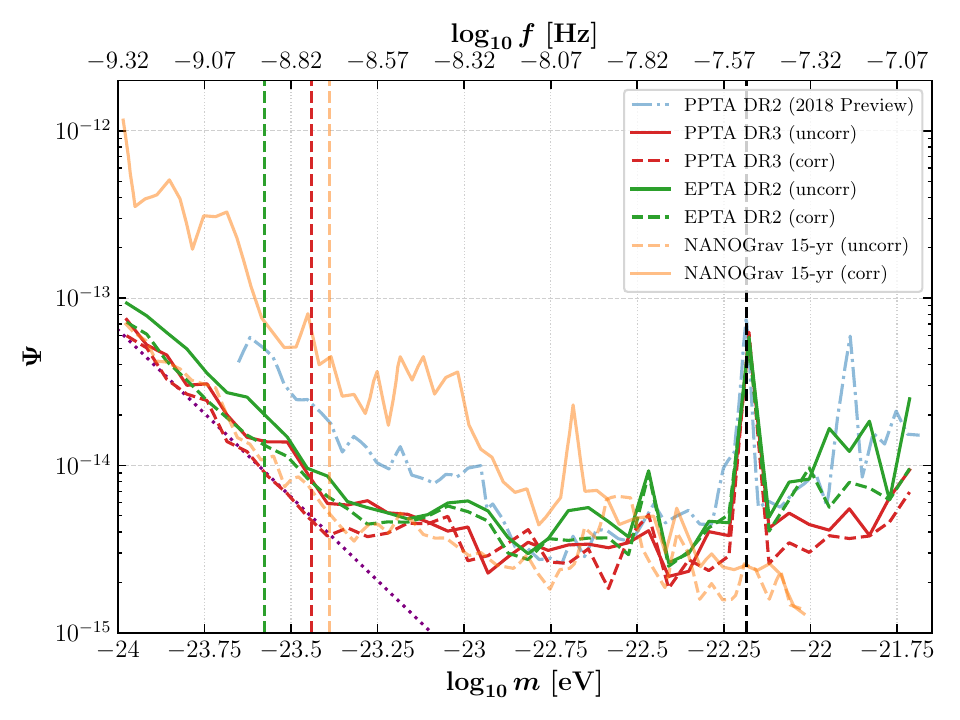}
\caption{Upper limits on scalar ULDM parameters from the analyses of the PPTA-DR3 (red) and EPTA-DR2 (green). The left panel shows constraints on the local dark matter density, while the right panel shows the corresponding limits on the dimensionless oscillation amplitude. Solid and dashed curves represent the uncorrelated and correlated cases, respectively. The purple dotted curve indicates the reference values assuming the local dark matter density of $\rho = 0.4\ \mrm{GeV/cm^3}$. Also shown for comparison are previous constraints from PPTA-DR2 (2018 Preview) (blue dot-dashed) \cite{2018_Porayko_PPTADM} and NANOGrav 15-yr (orange) \cite{2023_nanograv_newphysics}. Vertical dashed lines mark the lowest frequencies corresponding to the approximate observational timespans: PPTA-DR3 ($\sim$18 yr), EPTA-DR2 ({\tt{DR2full}}) ($\sim$24.7 yr), and NANOGrav 15-yr ($\sim$16 yr), with the black dashed line marking the one-over-one-year reference frequency. Note: The NANOGrav 15-yr analysis adopts $\log_{10}$-uniform prior for upper limits, and models the $\hat{\phi}$ parameters as independent in its correlated case.}
\label{scalar}
\eef

The 95\% confidence-level upper limits on the parameters of scalar ULDM are presented in \Fig{scalar}, showing constraints on the local dark matter density (left panel) and the oscillation amplitude (right panel). Our results do not rule out the scenario in which dark matter is entirely composed of scalar ULDM. The correlated signal model yields slightly more stringent constraints than its uncorrelated counterpart within the low mass range approximately $m \lesssim 7 \times 10^{-24}\ \mrm{eV}$. 
For comparison, we include previous constraints from PPTA-DR2 (2018 Preview) \cite{2018_Porayko_PPTADM}, as well as results from NANOGrav 15-yr \cite{2023_nanograv_newphysics}. The correlated constraints from PPTA-DR3 improve upon those from the PPTA-DR2 (2018 Preview) by approximately a factor of 5 for masses $ m \lesssim 5 \times 10^{-24} \ \eV$ and by a factor of about 2–5 over the intermediate range $5 \times 10^{-24} - 1 \times 10^{-23} \ \eV$. Within the range $ 1 \times 10^{-23} - 4 \times 10^{-23} \ \eV$, the constraints are broadly comparable between the two data releases. At higher masses $m \gtrsim 4 \times 10^{-23} \ \eV$, the constraints again improve, showing an enhancement by approximately a factor of 4 relative to the earlier results. The EPTA-DR2 limits are generally consistent with the PPTA-DR3 results. When compared to the limits from the NANOGrav 15-yr data, our constraints exhibit similar strength as in the uncorrelated case.

\bef[h]
\centering
\includegraphics[width=0.45\textwidth]{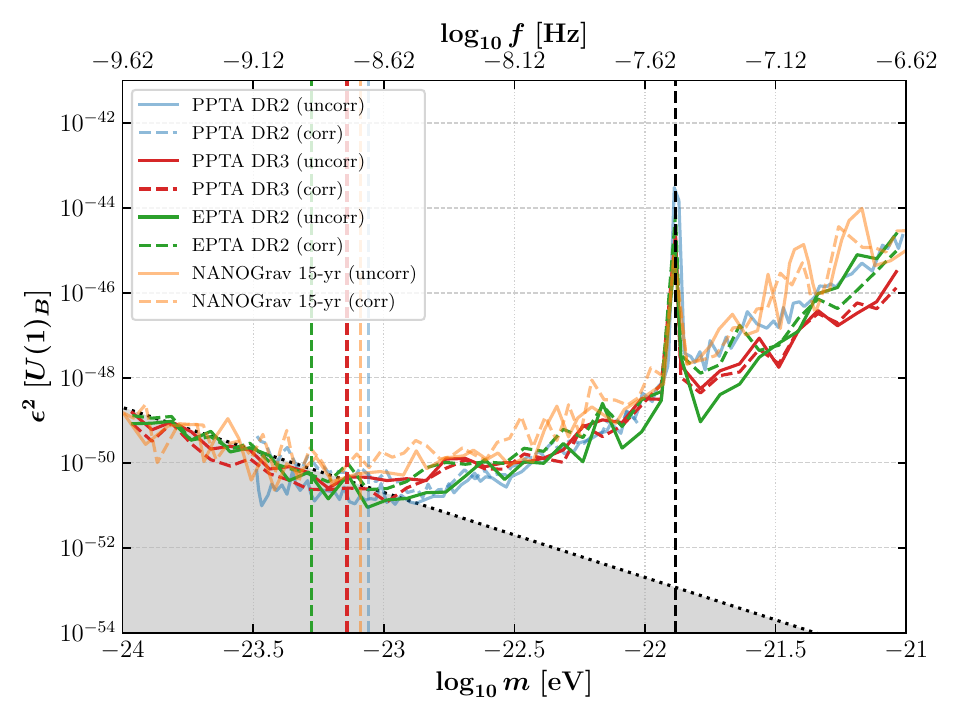}
\includegraphics[width=0.45\textwidth]{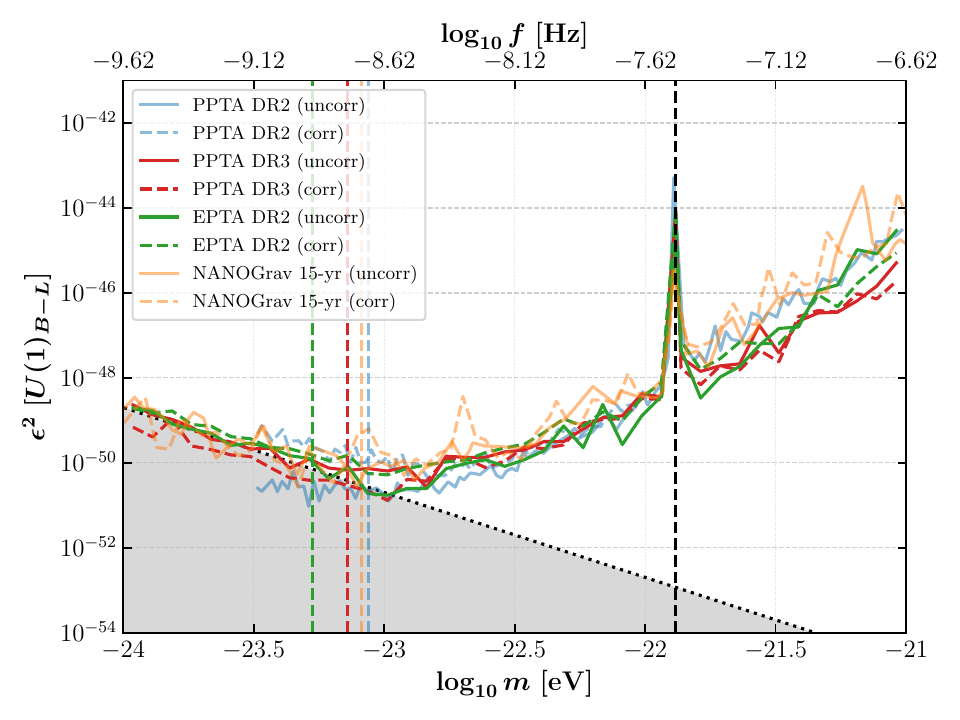}
\caption{Constraints on the dark photon coupling strength $\epsilon^2$ for the $U(1)_B$ (left) and $U(1)_{B-L}$ (right) interactions from this analysis, including results from PPTA-DR3 (red) and EPTA-DR2 (green). Solid and dashed curves represent the uncorrelated and correlated cases, respectively. The black dotted line and gray shaded region indicate the parameter space where gravitational effects dominate over fifth-force couplings, based on the simplified assumption that vector ULDM produces gravitational oscillations three times stronger than scalar ULDM. Comparisons are shown with previous PPTA-DR2 limits (blue) \cite{2018_PPTA_darkphoton_xuexiao} and NANOGrav 15-yr results (orange) \cite{2023_nanograv_newphysics}. Vertical dashed lines indicate the approximate observational timespans: PPTA-DR2 ($\sim$15 yr), PPTA-DR3 ($\sim$18 yr), EPTA-DR2 ({\tt{DR2full}}) ($\sim$24.7 yr), and NANOGrav 15-yr ($\sim$ 16 yr), with the black dashed line marking the one-over-one-year reference frequency. Note: The NANOGrav 15-yr analysis adopts $\log_{10}$-uniform prior for upper limits, and models the $\hat{\phi}$ parameters as independent in its correlated case.}
\label{dpdm}
\eef

We present the 95\% confidence upper limits on the DPDM coupling strength in \Fig{dpdm}, with the left and right panels corresponding to the $U(1)_{B}$ and $U(1)_{B-L}$ interactions, respectively. The gray shaded regions indicate the parameter space where gravitational effects dominate over fifth-force coupling. In the low-mass regime $m \lesssim 10^{-23} \ \eV$, gravitational signal constitute the dominant effect, leading to less reliable constraints in this region. Our results from EPTA-DR2 and PPTA-DR3 are comparable to previous constraints from NANOGrav 15-yr. 

\subsection{Discussion}

Our search for the gravitational signal of scalar ULDM does not exclude the possibility that ULDM constitutes all dark matter. Furthermore, the derived upper limits around $10^{-9} \ \Hz$ approach the reference value set by the local dark matter density. Future data releases, such as International Pulsar Timing Array’s third data release, may provide the sensitivity required to test this possibility conclusively. The differences between our EPTA-DR2 results and previous work \cite{2023_EPTA_uldm} stem from our adoption of customised pulsar noise model and linear-uniform prior for the upper limits. A detailed comparison of the results is presented in Appendix \ref{eptadr2_compare}. Accurate pulsar noise modeling is critical in PTA analyses. An imperfect model can bias the results by either absorbing a potential astrophysical signal or leaving residual noise unaccounted for, which can artificially weaken or strengthen the inferred upper limits. Recent studies employing hierarchical Bayesian frameworks confirm that the noise model significantly impacts the sensitivity and outcomes of GWB searches \cite{2024_vanHaasteren,2024_Goncharov}.

The impact of the dark matter density profile uncertainty on our final results is minor. This is because the results depend primarily on the ratio $\delta\rho = \rho(x_p)/\rho(x_e)$, which scales the ULDM signal strength. In the PPTA-DR3 and EPTA-DR2 dataset, all pulsars lie within about 1 kpc from Earth, where the dark matter density variation across different plausible profiles remains small. For a typical NFW profile, \(\delta\rho\) varies roughly between 0.8 and 1.6, the amplitude variation for the pulsar term is at most a factor of 1.6 (between 0.8 and 1.6). Consequently, the uncertainty in the DM profile propagates only to a modest scaling factor and, for the current sample, does not significantly affect our final upper limits.

The coherence of ULDM is important for its detection in PTA; therefore, measuring the correlation function is essential to verify the nature of any detected signal. Specifically, in the fast model the signal remains temporally deterministic but spatially stochastic \cite{2024_luuhoangnhan,2025_Boddy}. 
A key ongoing challenge in precision correlation modeling is the uncertainty in pulsar distances, which can be addressed by future improvements from VLBI \cite{2023_ding_vlbi} and, for some binary pulsars, by astrometric measurements from GAIA \cite{2023_Moran}.

For DPDM, our analysis assumes that DPDM constitutes the entirety of the dark matter content. If DPDM only comprises a subcomponent of the total dark matter abundance, the resulting constraints can be rescaled. Another important consideration is that gravitational effects dominate over fifth-force couplings in some of the mass range we have considered. Modeling both types of signals is further complicated in the case of vector ULDM, where the oscillation involves multiple components and uncertain directional properties \cite{2020_NomuraKimihiro_vector,2023_Omiya}. It is important to note the well-known degeneracy between the coupling strength and the local dark matter density in a combined signal fit \cite{2025_wu}. This correlation presents a common challenge for uniquely determining either parameter. Future searches could target the signal from vector ULDM oscillations using the correlation function \cite{2023_Omiya,2025_Dror}.


\subsection{Summary}
In this paper, we have searched for ULDM using the PPTA-DR3 and EPTA-DR2, targeting the gravitational signal of scalar ULDM and the fifth-force signal from DPDM. Expanding on previous work, we have introduced pulsar distances in the analysis to use a more physically motivated model for the ULDM signal. For scalar ULDM, our analysis does not exclude the scenario in which ULDM constitutes all of dark matter. Our upper limits improve upon those from the PPTA‑DR2 (2018 Preview) at lower masses, while remaining broadly comparable at higher masses. They are also consistent with the recent uncorrelated limits reported by the NANOGrav 15-yr. For DPDM, in the mass range $m \lesssim 1 \times 10^{-23} \ \eV$, gravitational effects dominate over fifth‑force couplings. Our constraints from the PPTA-DR3 and EPTA-DR2 are comparable with previous bounds and with the results from the NANOGrav 15-yr analysis.

\begin{acknowledgments}

X.J.Z is supported by the National Key Research and Development Program of China (No.~2023YFC2206704), the Fundamental Research Funds for the Central Universities, and the Supplemental Funds for Major Scientific Research Projects of Beijing Normal University (Zhuhai) under Project ZHPT2025001. H. H. acknowledges support from the PRIME programme of the German Academic Exchange Service (DAAD) with funds from the German Federal Ministry of Education and Research (BMBF). J.B.W is supported by Major Science and Technology Program of Xinjiang Uygur Autonomous Region, grant number 2022A03013-4. The work of C.S. was supported by the European Union – NextGeneration EU, mission 4, component 1, CUP C93C24004930006. Part of this work was supported by the ARC Centre of Excellence of Gravitational Wave Discovery (CE230100016). We also would greatly thank the referee for very useful comments.




Murriyang, the Parkes 64 m radio telescope is part of the Australia Telescope National Facility (\url{https://ror.org/05qajvd42}), which is funded by the Australian Government for operation as a National Facility managed by CSIRO. We acknowledge the Wiradjuri People as the Traditional Owners of the Observatory site. We acknowledge the Wurundjeri People of the Kulin Nation and the Wallumedegal People of the Darug Nation as the Traditional Owners of the land where this work was primarily carried out. We thank the Parkes Observatory staff for their support of the project for nearly two decades.

The European Pulsar Timing Array (EPTA) is a collaboration between
European and partner institutes, namely ASTRON (NL), INAF/Osservatorio
di Cagliari (IT), Max-Planck-Institut f\"{u}r Radioastronomie (GER),
Nan\c{c}ay/Paris Observatory (FRA), the University of Manchester (UK),
the University of Birmingham (UK), the University of East Anglia (UK),
the University of Bielefeld (GER), the University of Paris (FRA), the
University of Milan-Bicocca (IT), Peking University (CHN) and Shanghai Astronomical Observatory, CAS (CHN), with the
aim to provide high precision pulsar timing to work towards the direct
detection of low-frequency gravitational waves. An Advanced Grant of
the European Research Council to implement the Large European Array
for Pulsars (LEAP) has also provided funding. The EPTA is part of the
International Pulsar Timing Array (IPTA); we would like to thank our
IPTA colleagues for their help with this paper.

Part of this work is based on observations with the 100-m telescope of
the Max-Planck-Institut f\"{u}r Radioastronomie (MPIfR) at Effelsberg
in Germany. Pulsar research at the Jodrell Bank Centre for
Astrophysics and the observations using the Lovell Telescope are
supported by a Consolidated Grant (ST/T000414/1) from the UK's Science
and Technology Facilities Council (STFC). ICN is also supported by the
STFC doctoral training grant ST/T506291/1. The Nan{\c c}ay radio
Observatory is operated by the Paris Observatory, associated with the
French Centre National de la Recherche Scientifique (CNRS), and
partially supported by the Region Centre in France. We acknowledge
financial support from ``Programme National de Cosmologie and
Galaxies'' (PNCG), and ``Programme National Hautes Energies'' (PNHE)
funded by CNRS/INSU-IN2P3-INP, CEA and CNES, France. We acknowledge
financial support from Agence Nationale de la Recherche
(ANR-18-CE31-0015), France. The Westerbork Synthesis Radio Telescope
is operated by the Netherlands Institute for Radio Astronomy (ASTRON)
with support from the Netherlands Foundation for Scientific Research
(NWO). The Sardinia Radio Telescope (SRT) is funded by the Department
of University and Research (MIUR), the Italian Space Agency (ASI), and
the Autonomous Region of Sardinia (RAS) and is operated as a National
Facility by the National Institute for Astrophysics (INAF).

\end{acknowledgments}

\section*{Data Availability}
The data sets used for the findings of this article are openly available.
The PPTA-DR3 dataset can be accessed via the Parkes Pulsar Data Archive on the CSIRO Data Access Portal at (\url{http://data.csiro.au}). The corresponding customised pulsar noise models and analysis code are available at \url{https://github.com/danielreardon/PPTA-DR3}. The EPTA-DR2 dataset is available from \url{https://zenodo.org/records/8164425}, and its associated customised pulsar noise models and analysis are provided at \url{https://gitlab.in2p3.fr/epta/epta-dr2}. The signal model used in this paper is available upon request.

\bibliography{ref}

@ARTICLE{pptadr2_data,
       author = {{Kerr}, Matthew and {Reardon}, Daniel J. and {Hobbs}, George and {Shannon}, Ryan M. and {Manchester}, Richard N. and {Dai}, Shi and {Russell}, Christopher J. and {Zhang}, Songbo and {van Straten}, Willem and {Os{\l}owski}, Stefan and {Parthasarathy}, Aditya and {Spiewak}, Renee and {Bailes}, Matthew and {Bhat}, N.~D. Ramesh and {Cameron}, Andrew D. and {Coles}, William A. and {Dempsey}, James and {Deng}, Xinping and {Goncharov}, Boris and {Kaczmarek}, Jane F. and {Keith}, Michael J. and {Lasky}, Paul D. and {Lower}, Marcus E. and {Preisig}, Brett and {Sarkissian}, John Mihran and {Toomey}, Lawrence and {Wang}, Hongguang and {Wang}, Jingbo and {Zhang}, Lei and {Zhu}, Xingjiang},
        title = "{The Parkes Pulsar Timing Array project: second data release}",
      journal = {\pasa},
         year = 2020,
        month = jun,
       volume = {37},
          eid = {e020},
        pages = {e020},
          doi = {10.1017/pasa.2020.11},
archivePrefix = {arXiv},
       eprint = {2003.09780},
 primaryClass = {astro-ph.IM},
       adsurl = {https://ui.adsabs.harvard.edu/abs/2020PASA...37...20K}
}

@ARTICLE{pptadr3_data,
       author = {{Zic}, Andrew and {Reardon}, Daniel J. and {Kapur}, Agastya and {Hobbs}, George and {Mandow}, Rami and {Cury{\l}o}, Ma{\l}gorzata and {Shannon}, Ryan M. and {Askew}, Jacob and {Bailes}, Matthew and {Bhat}, N.~D. Ramesh and {Cameron}, Andrew and {Chen}, Zu-Cheng and {Dai}, Shi and {Di Marco}, Valentina and {Feng}, Yi and {Kerr}, Matthew and {Kulkarni}, Atharva and {Lower}, Marcus E. and {Luo}, Rui and {Manchester}, Richard N. and {Miles}, Matthew T. and {Nathan}, Rowina S. and {Os{\l}owski}, Stefan and {Rogers}, Axl F. and {Russell}, Christopher J. and {Sarkissian}, John M. and {Shamohammadi}, Mohsen and {Spiewak}, Ren{\'e}e and {Thyagarajan}, Nithyanandan and {Toomey}, Lawrence and {Wang}, Shuangqiang and {Zhang}, Lei and {Zhang}, Songbo and {Zhu}, Xing-Jiang},
        title = "{The Parkes Pulsar Timing Array third data release}",
      journal = {\pasa},
         year = 2023,
        month = dec,
       volume = {40},
          eid = {e049},
        pages = {e049},
          doi = {10.1017/pasa.2023.36},
archivePrefix = {arXiv},
       eprint = {2306.16230},
 primaryClass = {astro-ph.HE},
       adsurl = {https://ui.adsabs.harvard.edu/abs/2023PASA...40...49Z}
}

@ARTICLE{pptadr3_gwb,
       author = {{Reardon}, Daniel J. and {Zic}, Andrew and {Shannon}, Ryan M. and {Hobbs}, George B. and {Bailes}, Matthew and {Di Marco}, Valentina and {Kapur}, Agastya and {Rogers}, Axl F. and {Thrane}, Eric and {Askew}, Jacob and {Bhat}, N.~D. Ramesh and {Cameron}, Andrew and {Cury{\l}o}, Ma{\l}gorzata and {Coles}, William A. and {Dai}, Shi and {Goncharov}, Boris and {Kerr}, Matthew and {Kulkarni}, Atharva and {Levin}, Yuri and {Lower}, Marcus E. and {Manchester}, Richard N. and {Mandow}, Rami and {Miles}, Matthew T. and {Nathan}, Rowina S. and {Os{\l}owski}, Stefan and {Russell}, Christopher J. and {Spiewak}, Ren{\'e}e and {Zhang}, Songbo and {Zhu}, Xing-Jiang},
        title = "{Search for an Isotropic Gravitational-wave Background with the Parkes Pulsar Timing Array}",
      journal = {\apjl},
         year = 2023,
        month = jul,
       volume = {951},
       number = {1},
          eid = {L6},
        pages = {L6},
          doi = {10.3847/2041-8213/acdd02},
archivePrefix = {arXiv},
       eprint = {2306.16215},
 primaryClass = {astro-ph.HE},
       adsurl = {https://ui.adsabs.harvard.edu/abs/2023ApJ...951L...6R}
}

@ARTICLE{pptadr3_noise,
       author = {{Reardon}, Daniel J. and {Zic}, Andrew and {Shannon}, Ryan M. and {Di Marco}, Valentina and {Hobbs}, George B. and {Kapur}, Agastya and {Lower}, Marcus E. and {Mandow}, Rami and {Middleton}, Hannah and {Miles}, Matthew T. and {Rogers}, Axl F. and {Askew}, Jacob and {Bailes}, Matthew and {Bhat}, N.~D. Ramesh and {Cameron}, Andrew and {Kerr}, Matthew and {Kulkarni}, Atharva and {Manchester}, Richard N. and {Nathan}, Rowina S. and {Russell}, Christopher J. and {Os{\l}owski}, Stefan and {Zhu}, Xing-Jiang},
        title = "{The Gravitational-wave Background Null Hypothesis: Characterizing Noise in Millisecond Pulsar Arrival Times with the Parkes Pulsar Timing Array}",
      journal = {\apjl},
         year = 2023,
        month = jul,
       volume = {951},
       number = {1},
          eid = {L7},
        pages = {L7},
          doi = {10.3847/2041-8213/acdd03},
archivePrefix = {arXiv},
       eprint = {2306.16229},
 primaryClass = {astro-ph.HE},
       adsurl = {https://ui.adsabs.harvard.edu/abs/2023ApJ...951L...7R}
}

@ARTICLE{ppta_uwl,
       author = {{Hobbs}, George and {Manchester}, Richard N. and {Dunning}, Alex and {Jameson}, Andrew and {Roberts}, Paul and {George}, Daniel and {Green}, J.~A. and {Tuthill}, John and {Toomey}, Lawrence and {Kaczmarek}, Jane F. and {Mader}, Stacy and {Marquarding}, Malte and {Ahmed}, Azeem and {Amy}, Shaun W. and {Bailes}, Matthew and {Beresford}, Ron and {Bhat}, N.~D.~R. and {Bock}, Douglas C. -J. and {Bourne}, Michael and {Bowen}, Mark and {Brothers}, Michael and {Cameron}, Andrew D. and {Carretti}, Ettore and {Carter}, Nick and {Castillo}, Santy and {Chekkala}, Raji and {Cheng}, Wan and {Chung}, Yoon and {Craig}, Daniel A. and {Dai}, Shi and {Dawson}, Joanne and {Dempsey}, James and {Doherty}, Paul and {Dong}, Bin and {Edwards}, Philip and {Ergesh}, Tuohutinuer and {Gao}, Xuyang and {Han}, JinLin and {Hayman}, Douglas and {Indermuehle}, Balthasar and {Jeganathan}, Kanapathippillai and {Johnston}, Simon and {Kanoniuk}, Henry and {Kesteven}, Michael and {Kramer}, Michael and {Leach}, Mark and {Mcintyre}, Vince and {Moss}, Vanessa and {Os{\l}owski}, Stefan and {Phillips}, Chris and {Pope}, Nathan and {Preisig}, Brett and {Price}, Daniel and {Reeves}, Ken and {Reilly}, Les and {Reynolds}, John and {Robishaw}, Tim and {Roush}, Peter and {Ruckley}, Tim and {Sadler}, Elaine and {Sarkissian}, John and {Severs}, Sean and {Shannon}, Ryan and {Smart}, Ken and {Smith}, Malcolm and {Smith}, Stephanie and {Sobey}, Charlotte and {Staveley-Smith}, Lister and {Tzioumis}, Anastasios and {van Straten}, Willem and {Wang}, Nina and {Wen}, Linqing and {Whiting}, Matthew},
        title = "{An ultra-wide bandwidth (704 to 4 032 MHz) receiver for the Parkes radio telescope}",
      journal = {\pasa},
         year = 2020,
        month = apr,
       volume = {37},
          eid = {e012},
        pages = {e012},
          doi = {10.1017/pasa.2020.2},
archivePrefix = {arXiv},
       eprint = {1911.00656},
 primaryClass = {astro-ph.IM},
       adsurl = {https://ui.adsabs.harvard.edu/abs/2020PASA...37...12H}
}

@ARTICLE{pptadr3_uwl_timing,
       author = {{Cury{\l}o}, Ma{\l}gorzata and {Pennucci}, Timothy T. and {Bailes}, Matthew and {Bhat}, N.~D. Ramesh and {Cameron}, Andrew D. and {Dai}, Shi and {Hobbs}, George and {Kapur}, Agastya and {Manchester}, Richard N. and {Mandow}, Rami and {Miles}, Matthew T. and {Russell}, Christopher J. and {Reardon}, Daniel J. and {Shannon}, Ryan M. and {Spiewak}, Ren{\'e}e and {van Straten}, Willem and {Zhu}, Xing-Jiang and {Zic}, Andrew},
        title = "{Wide-band Timing of the Parkes Pulsar Timing Array UWL Data}",
      journal = {\apj},
         year = 2023,
        month = feb,
       volume = {944},
       number = {2},
          eid = {128},
        pages = {128},
          doi = {10.3847/1538-4357/aca535},
archivePrefix = {arXiv},
       eprint = {2211.12924},
 primaryClass = {astro-ph.HE},
       adsurl = {https://ui.adsabs.harvard.edu/abs/2023ApJ...944..128C}
}

@ARTICLE{2014_Andrei,
       author = {{Khmelnitsky}, Andrei and {Rubakov}, Valery},
        title = "{Pulsar timing signal from ultralight scalar dark matter}",
      journal = {\jcap},
         year = 2014,
        month = feb,
       volume = {2014},
       number = {2},
          eid = {019},
        pages = {019},
          doi = {10.1088/1475-7516/2014/02/019},
archivePrefix = {arXiv},
       eprint = {1309.5888},
 primaryClass = {astro-ph.CO},
       adsurl = {https://ui.adsabs.harvard.edu/abs/2014JCAP...02..019K}
}

@ARTICLE{1996_profile,
       author = {{Navarro}, Julio F. and {Frenk}, Carlos S. and {White}, Simon D.~M.},
        title = "{The Structure of Cold Dark Matter Halos}",
      journal = {\apj},
         year = 1996,
        month = may,
       volume = {462},
        pages = {563},
          doi = {10.1086/177173},
archivePrefix = {arXiv},
       eprint = {astro-ph/9508025},
 primaryClass = {astro-ph},
       adsurl = {https://ui.adsabs.harvard.edu/abs/1996ApJ...462..563N}
}

@ARTICLE{2012_bovy,
       author = {{Bovy}, Jo and {Tremaine}, Scott},
        title = "{On the Local Dark Matter Density}",
      journal = {\apj},
         year = 2012,
        month = sep,
       volume = {756},
       number = {1},
          eid = {89},
        pages = {89},
          doi = {10.1088/0004-637X/756/1/89},
archivePrefix = {arXiv},
       eprint = {1205.4033},
 primaryClass = {astro-ph.GA},
       adsurl = {https://ui.adsabs.harvard.edu/abs/2012ApJ...756...89B}
}

@ARTICLE{2014_Read,
    author = "Read, J. I.",
    title = "{The Local Dark Matter Density}",
    eprint = "1404.1938",
    archivePrefix = "arXiv",
    primaryClass = "astro-ph.GA",
    reportNumber = "JPHYSG-100038.R1",
    doi = "10.1088/0954-3899/41/6/063101",
    journal = "J. Phys. G",
    volume = "41",
    pages = "063101",
    year = "2014"
}

@ARTICLE{2018_Sivertsson,
       author = {{Sivertsson}, S. and {Silverwood}, H. and {Read}, J.~I. and {Bertone}, G. and {Steger}, P.},
        title = "{The local dark matter density from SDSS-SEGUE G-dwarfs}",
      journal = {\mnras},
         year = 2018,
        month = aug,
       volume = {478},
       number = {2},
        pages = {1677-1693},
          doi = {10.1093/mnras/sty977},
archivePrefix = {arXiv},
       eprint = {1708.07836},
 primaryClass = {astro-ph.GA},
       adsurl = {https://ui.adsabs.harvard.edu/abs/2018MNRAS.478.1677S}
}

@ARTICLE{2020_desalas,
    author = "de Salas, Pablo F.",
    editor = "Nakahata, Masayuki",
    title = "{Dark matter local density determination based on recent observations}",
    eprint = "1910.14366",
    archivePrefix = "arXiv",
    primaryClass = "astro-ph.GA",
    doi = "10.1088/1742-6596/1468/1/012020",
    journal = "J. Phys. Conf. Ser.",
    volume = "1468",
    number = "1",
    pages = "012020",
    year = "2020"
}

@ARTICLE{2014_Porayko,
       author = {{Porayko}, N.~K. and {Postnov}, K.~A.},
        title = "{Constraints on ultralight scalar dark matter from pulsar timing}",
      journal = {\prd},
         year = 2014,
        month = sep,
       volume = {90},
       number = {6},
          eid = {062008},
        pages = {062008},
          doi = {10.1103/PhysRevD.90.062008},
archivePrefix = {arXiv},
       eprint = {1408.4670},
 primaryClass = {astro-ph.CO},
       adsurl = {https://ui.adsabs.harvard.edu/abs/2014PhRvD..90f2008P}
}

@ARTICLE{2018_Porayko_PPTADM,
       author = {{Porayko}, Nataliya K. and {Zhu}, Xingjiang and {Levin}, Yuri and {Hui}, Lam and {Hobbs}, George and {Grudskaya}, Aleksandra and {Postnov}, Konstantin and {Bailes}, Matthew and {Bhat}, N.~D. Ramesh and {Coles}, William and {Dai}, Shi and {Dempsey}, James and {Keith}, Michael J. and {Kerr}, Matthew and {Kramer}, Michael and {Lasky}, Paul D. and {Manchester}, Richard N. and {Os{\l}owski}, Stefan and {Parthasarathy}, Aditya and {Ravi}, Vikram and {Reardon}, Daniel J. and {Rosado}, Pablo A. and {Russell}, Christopher J. and {Shannon}, Ryan M. and {Spiewak}, Ren{\'e}e and {van Straten}, Willem and {Toomey}, Lawrence and {Wang}, Jingbo and {Wen}, Linqing and {You}, Xiaopeng and {PPTA Collaboration}},
        title = "{Parkes Pulsar Timing Array constraints on ultralight scalar-field dark matter}",
      journal = {\prd},
         year = 2018,
        month = nov,
       volume = {98},
       number = {10},
          eid = {102002},
        pages = {102002},
          doi = {10.1103/PhysRevD.98.102002},
archivePrefix = {arXiv},
       eprint = {1810.03227},
 primaryClass = {astro-ph.CO},
       adsurl = {https://ui.adsabs.harvard.edu/abs/2018PhRvD..98j2002P}
}

@ARTICLE{2023_kim,
       author = {{Kim}, Hyungjin},
        title = "{Gravitational interaction of ultralight dark matter with interferometers}",
      journal = {\jcap},
         year = 2023,
        month = dec,
       volume = {2023},
       number = {12},
          eid = {018},
        pages = {018},
          doi = {10.1088/1475-7516/2023/12/018},
archivePrefix = {arXiv},
       eprint = {2306.13348},
 primaryClass = {hep-ph},
       adsurl = {https://ui.adsabs.harvard.edu/abs/2023JCAP...12..018K}
}

@ARTICLE{2024_Kim_pta,
       author = {{Kim}, Hyungjin and {Mitridate}, Andrea},
        title = "{Stochastic ultralight dark matter fluctuations in pulsar timing arrays}",
      journal = {\prd},
         year = 2024,
        month = mar,
       volume = {109},
       number = {5},
          eid = {055017},
        pages = {055017},
          doi = {10.1103/PhysRevD.109.055017},
archivePrefix = {arXiv},
       eprint = {2312.12225},
 primaryClass = {hep-ph},
       adsurl = {https://ui.adsabs.harvard.edu/abs/2024PhRvD.109e5017K}
}

@ARTICLE{12.5cw_nanograv,
       author = {{Arzoumanian}, Zaven and {Baker}, Paul T. and {Blecha}, Laura and {Blumer}, Harsha and {Brazier}, Adam and {Brook}, Paul R. and {Burke-Spolaor}, Sarah and {B{\'e}csy}, Bence and {Casey-Clyde}, J. Andrew and {Charisi}, Maria and {Chatterjee}, Shami and {Chen}, Siyuan and {Cordes}, James M. and {Cornish}, Neil J. and {Crawford}, Fronefield and {Cromartie}, H. Thankful and {Decesar}, Megan E. and {Demorest}, Paul B. and {Dolch}, Timothy and {Drachler}, Brendan and {Ellis}, Justin A. and {Ferrara}, E.~C. and {Fiore}, William and {Fonseca}, Emmanuel and {Freedman}, Gabriel E. and {Garver-Daniels}, Nathan and {Gentile}, Peter A. and {Glaser}, Joseph and {Good}, Deborah C. and {G{\"u}ltekin}, Kayhan and {Hazboun}, Jeffrey S. and {Jennings}, Ross J. and {Johnson}, Aaron D. and {Jones}, Megan L. and {Kaiser}, Andrew R. and {Kaplan}, David L. and {Kelley}, Luke Zoltan and {Key}, Joey Shapiro and {Laal}, Nima and {Lam}, Michael T. and {Lamb}, William G. and {W. Lazio}, T. Joseph and {Lewandowska}, Natalia and {Liu}, Tingting and {Lorimer}, Duncan R. and {Luo}, Jing and {Lynch}, Ryan S. and {Madison}, Dustin R. and {McEwen}, Alexander and {McLaughlin}, Maura A. and {Mingarelli}, Chiara M.~F. and {Ng}, Cherry and {Nice}, David J. and {Ocker}, Stella Koch and {Olum}, Ken D. and {Pennucci}, Timothy T. and {Pol}, Nihan S. and {Ransom}, Scott M. and {Ray}, Paul S. and {Romano}, Joseph D. and {Shapiro-Albert}, Brent J. and {Siemens}, Xavier and {Simon}, Joseph and {Siwek}, Magdalena and {Spiewak}, Ren{\'e}e and {Stairs}, Ingrid H. and {Stinebring}, Daniel R. and {Stovall}, Kevin and {Swiggum}, Joseph K. and {Sydnor}, Jessica and {Taylor}, Stephen R. and {Turner}, Jacob E. and {Vallisneri}, Michele and {Vigeland}, Sarah J. and {Wahl}, Haley M. and {Walsh}, Gregory and {Witt}, Caitlin A. and {Young}, Olivia and {Nanograv Collaboration}},
        title = "{The NANOGrav 12.5 yr Data Set: Bayesian Limits on Gravitational Waves from Individual Supermassive Black Hole Binaries}",
      journal = {\apjl},
         year = 2023,
        month = jul,
       volume = {951},
       number = {2},
          eid = {L28},
        pages = {L28},
          doi = {10.3847/2041-8213/acdbc7},
archivePrefix = {arXiv},
       eprint = {2301.03608},
 primaryClass = {astro-ph.GA},
       adsurl = {https://ui.adsabs.harvard.edu/abs/2023ApJ...951L..28A}
}

@ARTICLE{15cw_nanograv,
       author = {{Agazie}, Gabriella and {Anumarlapudi}, Akash and {Archibald}, Anne M. and {Arzoumanian}, Zaven and {Baker}, Paul T. and {B{\'e}csy}, Bence and {Blecha}, Laura and {Brazier}, Adam and {Brook}, Paul R. and {Burke-Spolaor}, Sarah and {Case}, Robin and {Casey-Clyde}, J. Andrew and {Charisi}, Maria and {Chatterjee}, Shami and {Cohen}, Tyler and {Cordes}, James M. and {Cornish}, Neil J. and {Crawford}, Fronefield and {Cromartie}, H. Thankful and {Crowter}, Kathryn and {Decesar}, Megan E. and {Demorest}, Paul B. and {Digman}, Matthew C. and {Dolch}, Timothy and {Drachler}, Brendan and {Ferrara}, Elizabeth C. and {Fiore}, William and {Fonseca}, Emmanuel and {Freedman}, Gabriel E. and {Garver-Daniels}, Nate and {Gentile}, Peter A. and {Glaser}, Joseph and {Good}, Deborah C. and {G{\"u}ltekin}, Kayhan and {Hazboun}, Jeffrey S. and {Hourihane}, Sophie and {Jennings}, Ross J. and {Johnson}, Aaron D. and {Jones}, Megan L. and {Kaiser}, Andrew R. and {Kaplan}, David L. and {Kelley}, Luke Zoltan and {Kerr}, Matthew and {Key}, Joey S. and {Laal}, Nima and {Lam}, Michael T. and {Lamb}, William G. and {Lazio}, T. Joseph W. and {Lewandowska}, Natalia and {Liu}, Tingting and {Lorimer}, Duncan R. and {Luo}, Jing and {Lynch}, Ryan S. and {Ma}, Chung-Pei and {Madison}, Dustin R. and {McEwen}, Alexander and {McKee}, James W. and {McLaughlin}, Maura A. and {McMann}, Natasha and {Meyers}, Bradley W. and {Meyers}, Patrick M. and {Mingarelli}, Chiara M.~F. and {Mitridate}, Andrea and {Ng}, Cherry and {Nice}, David J. and {Ocker}, Stella Koch and {Olum}, Ken D. and {Pennucci}, Timothy T. and {Perera}, Benetge B.~P. and {Petrov}, Polina and {Pol}, Nihan S. and {Radovan}, Henri A. and {Ransom}, Scott M. and {Ray}, Paul S. and {Romano}, Joseph D. and {Sardesai}, Shashwat C. and {Schmiedekamp}, Ann and {Schmiedekamp}, Carl and {Schmitz}, Kai and {Shapiro-Albert}, Brent J. and {Siemens}, Xavier and {Simon}, Joseph and {Siwek}, Magdalena S. and {Stairs}, Ingrid H. and {Stinebring}, Daniel R. and {Stovall}, Kevin and {Susobhanan}, Abhimanyu and {Swiggum}, Joseph K. and {Taylor}, Jacob and {Taylor}, Stephen R. and {Turner}, Jacob E. and {Unal}, Caner and {Vallisneri}, Michele and {van Haasteren}, Rutger and {Vigeland}, Sarah J. and {Wahl}, Haley M. and {Witt}, Caitlin A. and {Young}, Olivia and {Nanograv Collaboration}},
        title = "{The NANOGrav 15 yr Data Set: Bayesian Limits on Gravitational Waves from Individual Supermassive Black Hole Binaries}",
      journal = {\apjl},
         year = 2023,
        month = jul,
       volume = {951},
       number = {2},
          eid = {L50},
        pages = {L50},
          doi = {10.3847/2041-8213/ace18a},
archivePrefix = {arXiv},
       eprint = {2306.16222},
 primaryClass = {astro-ph.HE},
       adsurl = {https://ui.adsabs.harvard.edu/abs/2023ApJ...951L..50A}
}

@ARTICLE{2018_PPTA_darkphoton_xuexiao,
       author = {{Xue}, Xiao and {Xia}, Zi-Qing and {Zhu}, Xingjiang and {Zhao}, Yue and {Shu}, Jing and {Yuan}, Qiang and {Bhat}, N.~D. Ramesh and {Cameron}, Andrew D. and {Dai}, Shi and {Feng}, Yi and {Goncharov}, Boris and {Hobbs}, George and {Howard}, Eric and {Manchester}, Richard N. and {Parthasarathy}, Aditya and {Reardon}, Daniel J. and {Russell}, Christopher J. and {Shannon}, Ryan M. and {Spiewak}, Ren{\'e}e and {Thyagarajan}, Nithyanandan and {Wang}, Jingbo and {Zhang}, Lei and {Zhang}, Songbo and {PPTA Collaboration}},
        title = "{High-precision search for dark photon dark matter with the Parkes Pulsar Timing Array}",
      journal = {Phys. Rev. Res.},
         year = 2022,
        month = feb,
       volume = {4},
       number = {1},
          eid = {L012022},
        pages = {L012022},
          doi = {10.1103/PhysRevResearch.4.L012022},
archivePrefix = {arXiv},
       eprint = {2112.07687},
 primaryClass = {hep-ph},
       adsurl = {https://ui.adsabs.harvard.edu/abs/2022PhRvR...4a2022X}
}

@ARTICLE{2022_PPTA_vector_wuyumei,
       author = {{Wu}, Yu-Mei and {Chen}, Zu-Cheng and {Huang}, Qing-Guo and {Zhu}, Xingjiang and {Bhat}, N.~D. Ramesh and {Feng}, Yi and {Hobbs}, George and {Manchester}, Richard N. and {Russell}, Christopher J. and {Shannon}, R.~M. and {PPTA Collaboration}},
        title = "{Constraining ultralight vector dark matter with the Parkes Pulsar Timing Array second data release}",
      journal = {\prd},
         year = 2022,
        month = oct,
       volume = {106},
       number = {8},
          eid = {L081101},
        pages = {L081101},
          doi = {10.1103/PhysRevD.106.L081101},
archivePrefix = {arXiv},
       eprint = {2210.03880},
 primaryClass = {astro-ph.CO},
       adsurl = {https://ui.adsabs.harvard.edu/abs/2022PhRvD.106h1101W}
}

@ARTICLE{2023_EPTA_uldm,
       author = {{Smarra}, Clemente and {Goncharov}, Boris and {Barausse}, Enrico and {Antoniadis}, J. and {Babak}, S. and {Nielsen}, A. -S. Bak and {Bassa}, C.~G. and {Berthereau}, A. and {Bonetti}, M. and {Bortolas}, E. and {Brook}, P.~R. and {Burgay}, M. and {Caballero}, R.~N. and {Chalumeau}, A. and {Champion}, D.~J. and {Chanlaridis}, S. and {Chen}, S. and {Cognard}, I. and {Desvignes}, G. and {Falxa}, M. and {Ferdman}, R.~D. and {Franchini}, A. and {Gair}, J.~R. and {Graikou}, E. and {Grie{\ss}meier}, J. -M. and {Guillemot}, L. and {Guo}, Y.~J. and {Hu}, H. and {Iraci}, F. and {Izquierdo-Villalba}, D. and {Jang}, J. and {Jawor}, J. and {Janssen}, G.~H. and {Jessner}, A. and {Karuppusamy}, R. and {Keane}, E.~F. and {Keith}, M.~J. and {Kramer}, M. and {Krishnakumar}, M.~A. and {Lackeos}, K. and {Lee}, K.~J. and {Liu}, K. and {Liu}, Y. and {Lyne}, A.~G. and {McKee}, J.~W. and {Main}, R.~A. and {Mickaliger}, M.~B. and {Ni{\r{A}}{\textsterling}u}, I.~C. and {Parthasarathy}, A. and {Perera}, B.~B.~P. and {Perrodin}, D. and {Petiteau}, A. and {Porayko}, N.~K. and {Possenti}, A. and {Leclere}, H. Quelquejay and {Samajdar}, A. and {Sanidas}, S.~A. and {Sesana}, A. and {Shaifullah}, G. and {Speri}, L. and {Spiewak}, R. and {Stappers}, B.~W. and {Susarla}, S.~C. and {Theureau}, G. and {Tiburzi}, C. and {van der Wateren}, E. and {Vecchio}, A. and {Krishnan}, V. Venkatraman and {Wang}, J. and {Wang}, L. and {Wu}, Z. and {European Pulsar Timing Array}},
        title = "{Second Data Release from the European Pulsar Timing Array: Challenging the Ultralight Dark Matter Paradigm}",
      journal = {\prl},
         year = 2023,
        month = oct,
       volume = {131},
       number = {17},
          eid = {171001},
        pages = {171001},
          doi = {10.1103/PhysRevLett.131.171001},
archivePrefix = {arXiv},
       eprint = {2306.16228},
 primaryClass = {astro-ph.HE},
       adsurl = {https://ui.adsabs.harvard.edu/abs/2023PhRvL.131q1001S}
}

@ARTICLE{2023_nanograv_newphysics,
       author = {{Afzal}, Adeela and {Agazie}, Gabriella and {Anumarlapudi}, Akash and {Archibald}, Anne M. and {Arzoumanian}, Zaven and {Baker}, Paul T. and {B{\'e}csy}, Bence and {Blanco-Pillado}, Jose Juan and {Blecha}, Laura and {Boddy}, Kimberly K. and {Brazier}, Adam and {Brook}, Paul R. and {Burke-Spolaor}, Sarah and {Burnette}, Rand and {Case}, Robin and {Charisi}, Maria and {Chatterjee}, Shami and {Chatziioannou}, Katerina and {Cheeseboro}, Belinda D. and {Chen}, Siyuan and {Cohen}, Tyler and {Cordes}, James M. and {Cornish}, Neil J. and {Crawford}, Fronefield and {Cromartie}, H. Thankful and {Crowter}, Kathryn and {Cutler}, Curt J. and {Decesar}, Megan E. and {Degan}, Dallas and {Demorest}, Paul B. and {Deng}, Heling and {Dolch}, Timothy and {Drachler}, Brendan and {von Eckardstein}, Richard and {Ferrara}, Elizabeth C. and {Fiore}, William and {Fonseca}, Emmanuel and {Freedman}, Gabriel E. and {Garver-Daniels}, Nate and {Gentile}, Peter A. and {Gersbach}, Kyle A. and {Glaser}, Joseph and {Good}, Deborah C. and {Guertin}, Lydia and {G{\"u}ltekin}, Kayhan and {Hazboun}, Jeffrey S. and {Hourihane}, Sophie and {Islo}, Kristina and {Jennings}, Ross J. and {Johnson}, Aaron D. and {Jones}, Megan L. and {Kaiser}, Andrew R. and {Kaplan}, David L. and {Kelley}, Luke Zoltan and {Kerr}, Matthew and {Key}, Joey S. and {Laal}, Nima and {Lam}, Michael T. and {Lamb}, William G. and {Lazio}, T. Joseph W. and {Lee}, Vincent S.~H. and {Lewandowska}, Natalia and {Lino Dos Santos}, Rafael R. and {Littenberg}, Tyson B. and {Liu}, Tingting and {Lorimer}, Duncan R. and {Luo}, Jing and {Lynch}, Ryan S. and {Ma}, Chung-Pei and {Madison}, Dustin R. and {McEwen}, Alexander and {McKee}, James W. and {McLaughlin}, Maura A. and {McMann}, Natasha and {Meyers}, Bradley W. and {Meyers}, Patrick M. and {Mingarelli}, Chiara M.~F. and {Mitridate}, Andrea and {Nay}, Jonathan and {Natarajan}, Priyamvada and {Ng}, Cherry and {Nice}, David J. and {Ocker}, Stella Koch and {Olum}, Ken D. and {Pennucci}, Timothy T. and {Perera}, Benetge B.~P. and {Petrov}, Polina and {Pol}, Nihan S. and {Radovan}, Henri A. and {Ransom}, Scott M. and {Ray}, Paul S. and {Romano}, Joseph D. and {Sardesai}, Shashwat C. and {Schmiedekamp}, Ann and {Schmiedekamp}, Carl and {Schmitz}, Kai and {Schr{\"o}der}, Tobias and {Schult}, Levi and {Shapiro-Albert}, Brent J. and {Siemens}, Xavier and {Simon}, Joseph and {Siwek}, Magdalena S. and {Stairs}, Ingrid H. and {Stinebring}, Daniel R. and {Stovall}, Kevin and {Stratmann}, Peter and {Sun}, Jerry P. and {Susobhanan}, Abhimanyu and {Swiggum}, Joseph K. and {Taylor}, Jacob and {Taylor}, Stephen R. and {Trickle}, Tanner and {Turner}, Jacob E. and {Unal}, Caner and {Vallisneri}, Michele and {Verma}, Sonali and {Vigeland}, Sarah J. and {Wahl}, Haley M. and {Wang}, Qiaohong and {Witt}, Caitlin A. and {Wright}, David and {Young}, Olivia and {Zurek}, Kathryn M. and {NANOGrav Collaboration}},
        title = "{The NANOGrav 15 yr Data Set: Search for Signals from New Physics}",
      journal = {\apjl},
         year = 2023,
        month = jul,
       volume = {951},
       number = {1},
          eid = {L11},
        pages = {L11},
          doi = {10.3847/2041-8213/acdc91},
archivePrefix = {arXiv},
       eprint = {2306.16219},
 primaryClass = {astro-ph.HE},
       adsurl = {https://ui.adsabs.harvard.edu/abs/2023ApJ...951L..11A}
}

@ARTICLE{2000_fdm_hu,
       author = {{Hu}, Wayne and {Barkana}, Rennan and {Gruzinov}, Andrei},
        title = "{Fuzzy Cold Dark Matter: The Wave Properties of Ultralight Particles}",
      journal = {\prl},
         year = 2000,
        month = aug,
       volume = {85},
       number = {6},
        pages = {1158-1161},
          doi = {10.1103/PhysRevLett.85.1158},
archivePrefix = {arXiv},
       eprint = {astro-ph/0003365},
 primaryClass = {astro-ph},
       adsurl = {https://ui.adsabs.harvard.edu/abs/2000PhRvL..85.1158H}
}

@ARTICLE{2023_xiaziqing_gammapTA,
       author = {{Xia}, Zi-Qing and {Tang}, Tian-Peng and {Huang}, Xiaoyuan and {Yuan}, Qiang and {Fan}, Yi-Zhong},
        title = "{Constraining ultralight dark matter using the Fermi-LAT pulsar timing array}",
      journal = {\prd},
         year = 2023,
        month = jun,
       volume = {107},
       number = {12},
          eid = {L121302},
        pages = {L121302},
          doi = {10.1103/PhysRevD.107.L121302},
archivePrefix = {arXiv},
       eprint = {2303.17545},
 primaryClass = {astro-ph.HE},
       adsurl = {https://ui.adsabs.harvard.edu/abs/2023PhRvD.107l1302X}
}

@ARTICLE{2024_luuhoangnhan,
       author = {{Luu}, Hoang Nhan and {Liu}, Tao and {Ren}, Jing and {Broadhurst}, Tom and {Yang}, Ruizhi and {Wang}, Jie-Shuang and {Xie}, Zhen},
        title = "{Stochastic Wave Dark Matter with Fermi-LAT {\ensuremath{\gamma}}-Ray Pulsar Timing Array}",
      journal = {\apjl},
         year = 2024,
        month = mar,
       volume = {963},
       number = {2},
          eid = {L46},
        pages = {L46},
          doi = {10.3847/2041-8213/ad2ae2},
archivePrefix = {arXiv},
       eprint = {2304.04735},
 primaryClass = {astro-ph.HE},
       adsurl = {https://ui.adsabs.harvard.edu/abs/2024ApJ...963L..46L}
}

@ARTICLE{2020_katoryo,
       author = {{Kato}, Ryo and {Soda}, Jiro},
        title = "{Search for ultralight scalar dark matter with NANOGrav pulsar timing arrays}",
      journal = {\jcap},
         year = 2020,
        month = sep,
       volume = {2020},
       number = {9},
          eid = {036},
        pages = {036},
          doi = {10.1088/1475-7516/2020/09/036},
archivePrefix = {arXiv},
       eprint = {1904.09143},
 primaryClass = {astro-ph.HE},
       adsurl = {https://ui.adsabs.harvard.edu/abs/2020JCAP...09..036K}
}

@ARTICLE{2020_NomuraKimihiro_vector,
       author = {{Nomura}, Kimihiro and {Ito}, Asuka and {Soda}, Jiro},
        title = "{Pulsar timing residual induced by ultralight vector dark matter}",
      journal = {\epjc},
         year = 2020,
        month = may,
       volume = {80},
       number = {5},
          eid = {419},
        pages = {419},
          doi = {10.1140/epjc/s10052-020-7990-y},
archivePrefix = {arXiv},
       eprint = {1912.10210},
 primaryClass = {gr-qc},
       adsurl = {https://ui.adsabs.harvard.edu/abs/2020EPJC...80..419N}
}

@ARTICLE{2022_kaplan_clock,
       author = {{Kaplan}, David E. and {Mitridate}, Andrea and {Trickle}, Tanner},
        title = "{Constraining fundamental constant variations from ultralight dark matter with pulsar timing arrays}",
      journal = {\prd},
         year = 2022,
        month = aug,
       volume = {106},
       number = {3},
          eid = {035032},
        pages = {035032},
          doi = {10.1103/PhysRevD.106.035032},
archivePrefix = {arXiv},
       eprint = {2205.06817},
 primaryClass = {hep-ph},
       adsurl = {https://ui.adsabs.harvard.edu/abs/2022PhRvD.106c5032K}
}

@ARTICLE{2018_zhangjiajun,
       author = {{Zhang}, Jiajun and {Sming Tsai}, Yue-Lin and {Kuo}, Jui-Lin and {Cheung}, Kingman and {Chu}, Ming-Chung},
        title = "{Ultralight Axion Dark Matter and Its Impact on Dark Halo Structure in N-body Simulations}",
      journal = {\apj},
         year = 2018,
        month = jan,
       volume = {853},
       number = {1},
          eid = {51},
        pages = {51},
          doi = {10.3847/1538-4357/aaa485},
archivePrefix = {arXiv},
       eprint = {1611.00892},
 primaryClass = {astro-ph.CO},
       adsurl = {https://ui.adsabs.harvard.edu/abs/2018ApJ...853...51Z}
}

@article{2015_MarshDavid,
    author = "Marsh, David J. E.",
    title = "{Axion Cosmology}",
    eprint = "1510.07633",
    archivePrefix = "arXiv",
    primaryClass = "astro-ph.CO",
    reportNumber = "KCL-PH-TH-2015-50",
    doi = "10.1016/j.physrep.2016.06.005",
    journal = "Phys. Rept.",
    volume = "643",
    pages = "1--79",
    year = "2016"
}

@ARTICLE{2023_liutao,
       author = {{Liu}, Tao and {Lou}, Xuzixiang and {Ren}, Jing},
        title = "{Pulsar Polarization Arrays}",
      journal = {\prl},
         year = 2023,
        month = mar,
       volume = {130},
       number = {12},
          eid = {121401},
        pages = {121401},
          doi = {10.1103/PhysRevLett.130.121401},
archivePrefix = {arXiv},
       eprint = {2111.10615},
 primaryClass = {astro-ph.HE},
       adsurl = {https://ui.adsabs.harvard.edu/abs/2023PhRvL.130l1401L}
}

@ARTICLE{2014_Schive,
       author = {{Schive}, Hsi-Yu and {Chiueh}, Tzihong and {Broadhurst}, Tom},
        title = "{Cosmic structure as the quantum interference of a coherent dark wave}",
      journal = {Nature Physics},
         year = 2014,
        month = jul,
       volume = {10},
       number = {7},
        pages = {496-499},
          doi = {10.1038/nphys2996},
archivePrefix = {arXiv},
       eprint = {1406.6586},
 primaryClass = {astro-ph.GA},
       adsurl = {https://ui.adsabs.harvard.edu/abs/2014NatPh..10..496S}
}

@ARTICLE{2013_Nesti,
       author = {{Nesti}, Fabrizio and {Salucci}, Paolo},
        title = "{The Dark Matter halo of the Milky Way, AD 2013}",
      journal = {\jcap},
         year = 2013,
        month = jul,
       volume = {2013},
       number = {7},
          eid = {016},
        pages = {016},
          doi = {10.1088/1475-7516/2013/07/016},
archivePrefix = {arXiv},
       eprint = {1304.5127},
 primaryClass = {astro-ph.GA},
       adsurl = {https://ui.adsabs.harvard.edu/abs/2013JCAP...07..016N}
}

@ARTICLE{2014_Schive_3dsimulations,
       author = {{Schive}, Hsi-Yu and {Liao}, Ming-Hsuan and {Woo}, Tak-Pong and {Wong}, Shing-Kwong and {Chiueh}, Tzihong and {Broadhurst}, Tom and {Hwang}, W. -Y. Pauchy},
        title = "{Understanding the Core-Halo Relation of Quantum Wave Dark Matter from 3D Simulations}",
      journal = {\prl},
         year = 2014,
        month = dec,
       volume = {113},
       number = {26},
          eid = {261302},
        pages = {261302},
          doi = {10.1103/PhysRevLett.113.261302},
archivePrefix = {arXiv},
       eprint = {1407.7762},
 primaryClass = {astro-ph.GA},
       adsurl = {https://ui.adsabs.harvard.edu/abs/2014PhRvL.113z1302S}
}

@ARTICLE{2018_planck,
       author = {{Planck Collaboration} and {Aghanim}, N. and {Akrami}, Y. and {Ashdown}, M. and {Aumont}, J. and {Baccigalupi}, C. and {Ballardini}, M. and {Banday}, A.~J. and {Barreiro}, R.~B. and {Bartolo}, N. and {Basak}, S. and {Battye}, R. and {Benabed}, K. and {Bernard}, J. -P. and {Bersanelli}, M. and {Bielewicz}, P. and {Bock}, J.~J. and {Bond}, J.~R. and {Borrill}, J. and {Bouchet}, F.~R. and {Boulanger}, F. and {Bucher}, M. and {Burigana}, C. and {Butler}, R.~C. and {Calabrese}, E. and {Cardoso}, J. -F. and {Carron}, J. and {Challinor}, A. and {Chiang}, H.~C. and {Chluba}, J. and {Colombo}, L.~P.~L. and {Combet}, C. and {Contreras}, D. and {Crill}, B.~P. and {Cuttaia}, F. and {de Bernardis}, P. and {de Zotti}, G. and {Delabrouille}, J. and {Delouis}, J. -M. and {Di Valentino}, E. and {Diego}, J.~M. and {Dor{\'e}}, O. and {Douspis}, M. and {Ducout}, A. and {Dupac}, X. and {Dusini}, S. and {Efstathiou}, G. and {Elsner}, F. and {En{\ss}lin}, T.~A. and {Eriksen}, H.~K. and {Fantaye}, Y. and {Farhang}, M. and {Fergusson}, J. and {Fernandez-Cobos}, R. and {Finelli}, F. and {Forastieri}, F. and {Frailis}, M. and {Fraisse}, A.~A. and {Franceschi}, E. and {Frolov}, A. and {Galeotta}, S. and {Galli}, S. and {Ganga}, K. and {G{\'e}nova-Santos}, R.~T. and {Gerbino}, M. and {Ghosh}, T. and {Gonz{\'a}lez-Nuevo}, J. and {G{\'o}rski}, K.~M. and {Gratton}, S. and {Gruppuso}, A. and {Gudmundsson}, J.~E. and {Hamann}, J. and {Handley}, W. and {Hansen}, F.~K. and {Herranz}, D. and {Hildebrandt}, S.~R. and {Hivon}, E. and {Huang}, Z. and {Jaffe}, A.~H. and {Jones}, W.~C. and {Karakci}, A. and {Keih{\"a}nen}, E. and {Keskitalo}, R. and {Kiiveri}, K. and {Kim}, J. and {Kisner}, T.~S. and {Knox}, L. and {Krachmalnicoff}, N. and {Kunz}, M. and {Kurki-Suonio}, H. and {Lagache}, G. and {Lamarre}, J. -M. and {Lasenby}, A. and {Lattanzi}, M. and {Lawrence}, C.~R. and {Le Jeune}, M. and {Lemos}, P. and {Lesgourgues}, J. and {Levrier}, F. and {Lewis}, A. and {Liguori}, M. and {Lilje}, P.~B. and {Lilley}, M. and {Lindholm}, V. and {L{\'o}pez-Caniego}, M. and {Lubin}, P.~M. and {Ma}, Y. -Z. and {Mac{\'\i}as-P{\'e}rez}, J.~F. and {Maggio}, G. and {Maino}, D. and {Mandolesi}, N. and {Mangilli}, A. and {Marcos-Caballero}, A. and {Maris}, M. and {Martin}, P.~G. and {Martinelli}, M. and {Mart{\'\i}nez-Gonz{\'a}lez}, E. and {Matarrese}, S. and {Mauri}, N. and {McEwen}, J.~D. and {Meinhold}, P.~R. and {Melchiorri}, A. and {Mennella}, A. and {Migliaccio}, M. and {Millea}, M. and {Mitra}, S. and {Miville-Desch{\^e}nes}, M. -A. and {Molinari}, D. and {Montier}, L. and {Morgante}, G. and {Moss}, A. and {Natoli}, P. and {N{\o}rgaard-Nielsen}, H.~U. and {Pagano}, L. and {Paoletti}, D. and {Partridge}, B. and {Patanchon}, G. and {Peiris}, H.~V. and {Perrotta}, F. and {Pettorino}, V. and {Piacentini}, F. and {Polastri}, L. and {Polenta}, G. and {Puget}, J. -L. and {Rachen}, J.~P. and {Reinecke}, M. and {Remazeilles}, M. and {Renzi}, A. and {Rocha}, G. and {Rosset}, C. and {Roudier}, G. and {Rubi{\~n}o-Mart{\'\i}n}, J.~A. and {Ruiz-Granados}, B. and {Salvati}, L. and {Sandri}, M. and {Savelainen}, M. and {Scott}, D. and {Shellard}, E.~P.~S. and {Sirignano}, C. and {Sirri}, G. and {Spencer}, L.~D. and {Sunyaev}, R. and {Suur-Uski}, A. -S. and {Tauber}, J.~A. and {Tavagnacco}, D. and {Tenti}, M. and {Toffolatti}, L. and {Tomasi}, M. and {Trombetti}, T. and {Valenziano}, L. and {Valiviita}, J. and {Van Tent}, B. and {Vibert}, L. and {Vielva}, P. and {Villa}, F. and {Vittorio}, N. and {Wandelt}, B.~D. and {Wehus}, I.~K. and {White}, M. and {White}, S.~D.~M. and {Zacchei}, A. and {Zonca}, A.},
        title = "{Planck 2018 results. VI. Cosmological parameters}",
      journal = {\aap},
         year = 2020,
        month = sep,
       volume = {641},
          eid = {A6},
        pages = {A6},
          doi = {10.1051/0004-6361/201833910},
archivePrefix = {arXiv},
       eprint = {1807.06209},
 primaryClass = {astro-ph.CO},
       adsurl = {https://ui.adsabs.harvard.edu/abs/2020A&A...641A...6P}
}

@ARTICLE{1999_Bahcall,
       author = {{Bahcall}, N.~A. and {Ostriker}, J.~P. and {Perlmutter}, S. and {Steinhardt}, P.~J.},
        title = "{The Cosmic Triangle: Revealing the State of the Universe}",
      journal = {Science},
         year = 1999,
        month = may,
       volume = {284},
        pages = {1481},
          doi = {10.1126/science.284.5419.1481},
archivePrefix = {arXiv},
       eprint = {astro-ph/9906463},
 primaryClass = {astro-ph},
       adsurl = {https://ui.adsabs.harvard.edu/abs/1999Sci...284.1481B}
}

@ARTICLE{1994_Flores,
       author = {{Flores}, Ricardo A. and {Primack}, Joel R.},
        title = "{Observational and Theoretical Constraints on Singular Dark Matter Halos}",
      journal = {\apjl},
         year = 1994,
        month = may,
       volume = {427},
        pages = {L1},
          doi = {10.1086/187350},
archivePrefix = {arXiv},
       eprint = {astro-ph/9402004},
 primaryClass = {astro-ph},
       adsurl = {https://ui.adsabs.harvard.edu/abs/1994ApJ...427L...1F}
}

@ARTICLE{1994_Moore,
       author = {{Moore}, Ben},
        title = "{Evidence against dissipation-less dark matter from observations of galaxy haloes}",
      journal = {\nat},
         year = 1994,
        month = aug,
       volume = {370},
       number = {6491},
        pages = {629-631},
          doi = {10.1038/370629a0},
       adsurl = {https://ui.adsabs.harvard.edu/abs/1994Natur.370..629M}
}

@ARTICLE{1999_Klypin,
       author = {{Klypin}, Anatoly and {Kravtsov}, Andrey V. and {Valenzuela}, Octavio and {Prada}, Francisco},
        title = "{Where Are the Missing Galactic Satellites?}",
      journal = {\apj},
         year = 1999,
        month = sep,
       volume = {522},
       number = {1},
        pages = {82-92},
          doi = {10.1086/307643},
archivePrefix = {arXiv},
       eprint = {astro-ph/9901240},
 primaryClass = {astro-ph},
       adsurl = {https://ui.adsabs.harvard.edu/abs/1999ApJ...522...82K}
}

@ARTICLE{1999_Moore,
       author = {{Moore}, Ben and {Ghigna}, Sebastiano and {Governato}, Fabio and {Lake}, George and {Quinn}, Thomas and {Stadel}, Joachim and {Tozzi}, Paolo},
        title = "{Dark Matter Substructure within Galactic Halos}",
      journal = {\apjl},
         year = 1999,
        month = oct,
       volume = {524},
       number = {1},
        pages = {L19-L22},
          doi = {10.1086/312287},
archivePrefix = {arXiv},
       eprint = {astro-ph/9907411},
 primaryClass = {astro-ph},
       adsurl = {https://ui.adsabs.harvard.edu/abs/1999ApJ...524L..19M}
}

@ARTICLE{Boylan-Kolchin:2011qkt,
    author = "Boylan-Kolchin, Michael and Bullock, James S. and Kaplinghat, Manoj",
    title = "{Too big to fail? The puzzling darkness of massive Milky Way subhaloes}",
    eprint = "1103.0007",
    archivePrefix = "arXiv",
    primaryClass = "astro-ph.CO",
    doi = "10.1111/j.1745-3933.2011.01074.x",
    journal = "Mon. Not. Roy. Astron. Soc.",
    volume = "415",
    pages = "L40",
    year = "2011"
}

@article{Bullock:2017xww,
    author = "Bullock, James S. and Boylan-Kolchin, Michael",
    title = "{Small-Scale Challenges to the $\Lambda$CDM Paradigm}",
    eprint = "1707.04256",
    archivePrefix = "arXiv",
    primaryClass = "astro-ph.CO",
    doi = "10.1146/annurev-astro-091916-055313",
    journal = "Ann. Rev. Astron. Astrophys.",
    volume = "55",
    pages = "343--387",
    year = "2017"
}

@ARTICLE{2015_Chan,
       author = {{Chan}, T.~K. and {Kere{\v{s}}}, D. and {O{\~n}orbe}, J. and {Hopkins}, P.~F. and {Muratov}, A.~L. and {Faucher-Gigu{\`e}re}, C. -A. and {Quataert}, E.},
        title = "{The impact of baryonic physics on the structure of dark matter haloes: the view from the FIRE cosmological simulations}",
      journal = {\mnras},
         year = 2015,
        month = dec,
       volume = {454},
       number = {3},
        pages = {2981-3001},
          doi = {10.1093/mnras/stv2165},
archivePrefix = {arXiv},
       eprint = {1507.02282},
 primaryClass = {astro-ph.GA},
       adsurl = {https://ui.adsabs.harvard.edu/abs/2015MNRAS.454.2981C}
}

@ARTICLE{2017_Hui,
       author = {{Hui}, Lam and {Ostriker}, Jeremiah P. and {Tremaine}, Scott and {Witten}, Edward},
        title = "{Ultralight scalars as cosmological dark matter}",
      journal = {\prd},
         year = 2017,
        month = feb,
       volume = {95},
       number = {4},
          eid = {043541},
        pages = {043541},
          doi = {10.1103/PhysRevD.95.043541},
archivePrefix = {arXiv},
       eprint = {1610.08297},
 primaryClass = {astro-ph.CO},
       adsurl = {https://ui.adsabs.harvard.edu/abs/2017PhRvD..95d3541H}
}

@ARTICLE{2015_Weinberg,
       author = {{Weinberg}, David H. and {Bullock}, James S. and {Governato}, Fabio and {Kuzio de Naray}, Rachel and {Peter}, Annika H.~G.},
        title = "{Cold dark matter: Controversies on small scales}",
      journal = {Proc. Nat. Acad. Sci.},
         year = 2015,
        month = oct,
       volume = {112},
       number = {40},
        pages = {12249-12255},
          doi = {10.1073/pnas.1308716112},
archivePrefix = {arXiv},
       eprint = {1306.0913},
 primaryClass = {astro-ph.CO},
       adsurl = {https://ui.adsabs.harvard.edu/abs/2015PNAS..11212249W}
}

@ARTICLE{2021_Ferreira,
       author = {{Ferreira}, Elisa G.~M.},
        title = "{Ultra-light dark matter}",
      journal = {\aapr},
         year = 2021,
        month = dec,
       volume = {29},
       number = {1},
          eid = {7},
        pages = {7},
          doi = {10.1007/s00159-021-00135-6},
archivePrefix = {arXiv},
       eprint = {2005.03254},
 primaryClass = {astro-ph.CO},
       adsurl = {https://ui.adsabs.harvard.edu/abs/2021A&ARv..29....7F}
}

@ARTICLE{2018_Pierce,
       author = {{Pierce}, Aaron and {Riles}, Keith and {Zhao}, Yue},
        title = "{Searching for Dark Photon Dark Matter with Gravitational-Wave Detectors}",
      journal = {\prl},
         year = 2018,
        month = aug,
       volume = {121},
       number = {6},
          eid = {061102},
        pages = {061102},
          doi = {10.1103/PhysRevLett.121.061102},
archivePrefix = {arXiv},
       eprint = {1801.10161},
 primaryClass = {hep-ph},
       adsurl = {https://ui.adsabs.harvard.edu/abs/2018PhRvL.121f1102P}
}

@article{2019_Centers,
    author = "Centers, Gary P. and others",
    title = "{Stochastic fluctuations of bosonic dark matter}",
    eprint = "1905.13650",
    archivePrefix = "arXiv",
    primaryClass = "astro-ph.CO",
    doi = "10.1038/s41467-021-27632-7",
    journal = "Nature Commun.",
    volume = "12",
    number = "1",
    pages = "7321",
    year = "2021"
}

@ARTICLE{2018_Foster,
       author = {{Foster}, Joshua W. and {Rodd}, Nicholas L. and {Safdi}, Benjamin R.},
        title = "{Revealing the dark matter halo with axion direct detection}",
      journal = {\prd},
         year = 2018,
        month = jun,
       volume = {97},
       number = {12},
          eid = {123006},
        pages = {123006},
          doi = {10.1103/PhysRevD.97.123006},
archivePrefix = {arXiv},
       eprint = {1711.10489},
 primaryClass = {astro-ph.CO},
       adsurl = {https://ui.adsabs.harvard.edu/abs/2018PhRvD..97l3006F}
}

@ARTICLE{2023_nanograv_gwb,
       author = {{Agazie}, Gabriella and {Anumarlapudi}, Akash and {Archibald}, Anne M. and {Arzoumanian}, Zaven and {Baker}, Paul T. and {B{\'e}csy}, Bence and {Blecha}, Laura and {Brazier}, Adam and {Brook}, Paul R. and {Burke-Spolaor}, Sarah and {Burnette}, Rand and {Case}, Robin and {Charisi}, Maria and {Chatterjee}, Shami and {Chatziioannou}, Katerina and {Cheeseboro}, Belinda D. and {Chen}, Siyuan and {Cohen}, Tyler and {Cordes}, James M. and {Cornish}, Neil J. and {Crawford}, Fronefield and {Cromartie}, H. Thankful and {Crowter}, Kathryn and {Cutler}, Curt J. and {Decesar}, Megan E. and {Degan}, Dallas and {Demorest}, Paul B. and {Deng}, Heling and {Dolch}, Timothy and {Drachler}, Brendan and {Ellis}, Justin A. and {Ferrara}, Elizabeth C. and {Fiore}, William and {Fonseca}, Emmanuel and {Freedman}, Gabriel E. and {Garver-Daniels}, Nate and {Gentile}, Peter A. and {Gersbach}, Kyle A. and {Glaser}, Joseph and {Good}, Deborah C. and {G{\"u}ltekin}, Kayhan and {Hazboun}, Jeffrey S. and {Hourihane}, Sophie and {Islo}, Kristina and {Jennings}, Ross J. and {Johnson}, Aaron D. and {Jones}, Megan L. and {Kaiser}, Andrew R. and {Kaplan}, David L. and {Kelley}, Luke Zoltan and {Kerr}, Matthew and {Key}, Joey S. and {Klein}, Tonia C. and {Laal}, Nima and {Lam}, Michael T. and {Lamb}, William G. and {Lazio}, T. Joseph W. and {Lewandowska}, Natalia and {Littenberg}, Tyson B. and {Liu}, Tingting and {Lommen}, Andrea and {Lorimer}, Duncan R. and {Luo}, Jing and {Lynch}, Ryan S. and {Ma}, Chung-Pei and {Madison}, Dustin R. and {Mattson}, Margaret A. and {McEwen}, Alexander and {McKee}, James W. and {McLaughlin}, Maura A. and {McMann}, Natasha and {Meyers}, Bradley W. and {Meyers}, Patrick M. and {Mingarelli}, Chiara M.~F. and {Mitridate}, Andrea and {Natarajan}, Priyamvada and {Ng}, Cherry and {Nice}, David J. and {Ocker}, Stella Koch and {Olum}, Ken D. and {Pennucci}, Timothy T. and {Perera}, Benetge B.~P. and {Petrov}, Polina and {Pol}, Nihan S. and {Radovan}, Henri A. and {Ransom}, Scott M. and {Ray}, Paul S. and {Romano}, Joseph D. and {Sardesai}, Shashwat C. and {Schmiedekamp}, Ann and {Schmiedekamp}, Carl and {Schmitz}, Kai and {Schult}, Levi and {Shapiro-Albert}, Brent J. and {Siemens}, Xavier and {Simon}, Joseph and {Siwek}, Magdalena S. and {Stairs}, Ingrid H. and {Stinebring}, Daniel R. and {Stovall}, Kevin and {Sun}, Jerry P. and {Susobhanan}, Abhimanyu and {Swiggum}, Joseph K. and {Taylor}, Jacob and {Taylor}, Stephen R. and {Turner}, Jacob E. and {Unal}, Caner and {Vallisneri}, Michele and {van Haasteren}, Rutger and {Vigeland}, Sarah J. and {Wahl}, Haley M. and {Wang}, Qiaohong and {Witt}, Caitlin A. and {Young}, Olivia and {Nanograv Collaboration}},
        title = "{The NANOGrav 15 yr Data Set: Evidence for a Gravitational-wave Background}",
      journal = {\apjl},
         year = 2023,
        month = jul,
       volume = {951},
       number = {1},
          eid = {L8},
        pages = {L8},
          doi = {10.3847/2041-8213/acdac6},
archivePrefix = {arXiv},
       eprint = {2306.16213},
 primaryClass = {astro-ph.HE},
       adsurl = {https://ui.adsabs.harvard.edu/abs/2023ApJ...951L...8A}
}

@ARTICLE{2023_epta_gwb,
       author = {{EPTA Collaboration} and {InPTA Collaboration} and {Antoniadis}, J. and {Arumugam}, P. and {Arumugam}, S. and {Babak}, S. and {Bagchi}, M. and {Bak Nielsen}, A. -S. and {Bassa}, C.~G. and {Bathula}, A. and {Berthereau}, A. and {Bonetti}, M. and {Bortolas}, E. and {Brook}, P.~R. and {Burgay}, M. and {Caballero}, R.~N. and {Chalumeau}, A. and {Champion}, D.~J. and {Chanlaridis}, S. and {Chen}, S. and {Cognard}, I. and {Dandapat}, S. and {Deb}, D. and {Desai}, S. and {Desvignes}, G. and {Dhanda-Batra}, N. and {Dwivedi}, C. and {Falxa}, M. and {Ferdman}, R.~D. and {Franchini}, A. and {Gair}, J.~R. and {Goncharov}, B. and {Gopakumar}, A. and {Graikou}, E. and {Grie{\ss}meier}, J. -M. and {Guillemot}, L. and {Guo}, Y.~J. and {Gupta}, Y. and {Hisano}, S. and {Hu}, H. and {Iraci}, F. and {Izquierdo-Villalba}, D. and {Jang}, J. and {Jawor}, J. and {Janssen}, G.~H. and {Jessner}, A. and {Joshi}, B.~C. and {Kareem}, F. and {Karuppusamy}, R. and {Keane}, E.~F. and {Keith}, M.~J. and {Kharbanda}, D. and {Kikunaga}, T. and {Kolhe}, N. and {Kramer}, M. and {Krishnakumar}, M.~A. and {Lackeos}, K. and {Lee}, K.~J. and {Liu}, K. and {Liu}, Y. and {Lyne}, A.~G. and {McKee}, J.~W. and {Maan}, Y. and {Main}, R.~A. and {Mickaliger}, M.~B. and {Ni{\c{t}}u}, I.~C. and {Nobleson}, K. and {Paladi}, A.~K. and {Parthasarathy}, A. and {Perera}, B.~B.~P. and {Perrodin}, D. and {Petiteau}, A. and {Porayko}, N.~K. and {Possenti}, A. and {Prabu}, T. and {Quelquejay Leclere}, H. and {Rana}, P. and {Samajdar}, A. and {Sanidas}, S.~A. and {Sesana}, A. and {Shaifullah}, G. and {Singha}, J. and {Speri}, L. and {Spiewak}, R. and {Srivastava}, A. and {Stappers}, B.~W. and {Surnis}, M. and {Susarla}, S.~C. and {Susobhanan}, A. and {Takahashi}, K. and {Tarafdar}, P. and {Theureau}, G. and {Tiburzi}, C. and {van der Wateren}, E. and {Vecchio}, A. and {Venkatraman Krishnan}, V. and {Verbiest}, J.~P.~W. and {Wang}, J. and {Wang}, L. and {Wu}, Z.},
        title = "{The second data release from the European Pulsar Timing Array. III. Search for gravitational wave signals}",
      journal = {\aap},
         year = 2023,
        month = oct,
       volume = {678},
          eid = {A50},
        pages = {A50},
          doi = {10.1051/0004-6361/202346844},
archivePrefix = {arXiv},
       eprint = {2306.16214},
 primaryClass = {astro-ph.HE},
       adsurl = {https://ui.adsabs.harvard.edu/abs/2023A&A...678A..50E}
}

@ARTICLE{2023_ppta_gwb,
       author = {{Reardon}, Daniel J. and {Zic}, Andrew and {Shannon}, Ryan M. and {Hobbs}, George B. and {Bailes}, Matthew and {Di Marco}, Valentina and {Kapur}, Agastya and {Rogers}, Axl F. and {Thrane}, Eric and {Askew}, Jacob and {Bhat}, N.~D. Ramesh and {Cameron}, Andrew and {Cury{\l}o}, Ma{\l}gorzata and {Coles}, William A. and {Dai}, Shi and {Goncharov}, Boris and {Kerr}, Matthew and {Kulkarni}, Atharva and {Levin}, Yuri and {Lower}, Marcus E. and {Manchester}, Richard N. and {Mandow}, Rami and {Miles}, Matthew T. and {Nathan}, Rowina S. and {Os{\l}owski}, Stefan and {Russell}, Christopher J. and {Spiewak}, Ren{\'e}e and {Zhang}, Songbo and {Zhu}, Xing-Jiang},
        title = "{Search for an Isotropic Gravitational-wave Background with the Parkes Pulsar Timing Array}",
      journal = {\apjl},
         year = 2023,
        month = jul,
       volume = {951},
       number = {1},
          eid = {L6},
        pages = {L6},
          doi = {10.3847/2041-8213/acdd02},
archivePrefix = {arXiv},
       eprint = {2306.16215},
 primaryClass = {astro-ph.HE},
       adsurl = {https://ui.adsabs.harvard.edu/abs/2023ApJ...951L...6R}
}

@ARTICLE{2023cpta,
       author = {{Xu}, Heng and {Chen}, Siyuan and {Guo}, Yanjun and {Jiang}, Jinchen and {Wang}, Bojun and {Xu}, Jiangwei and {Xue}, Zihan and {Caballero}, R. Nicolas and {Yuan}, Jianping and {Xu}, Yonghua and {Wang}, Jingbo and {Hao}, Longfei and {Luo}, Jingtao and {Lee}, Kejia and {Han}, Jinlin and {Jiang}, Peng and {Shen}, Zhiqiang and {Wang}, Min and {Wang}, Na and {Xu}, Renxin and {Wu}, Xiangping and {Manchester}, Richard and {Qian}, Lei and {Guan}, Xin and {Huang}, Menglin and {Sun}, Chun and {Zhu}, Yan},
        title = "{Searching for the Nano-Hertz Stochastic Gravitational Wave Background with the Chinese Pulsar Timing Array Data Release I}",
      journal = {Res. Astron. Astrophys.},
         year = 2023,
        month = jul,
       volume = {23},
       number = {7},
          eid = {075024},
        pages = {075024},
          doi = {10.1088/1674-4527/acdfa5},
archivePrefix = {arXiv},
       eprint = {2306.16216},
 primaryClass = {astro-ph.HE},
       adsurl = {https://ui.adsabs.harvard.edu/abs/2023RAA....23g5024X}
}

@ARTICLE{2017_taylor,
       author = {{Taylor}, S.~R. and {Lentati}, L. and {Babak}, S. and {Brem}, P. and {Gair}, J.~R. and {Sesana}, A. and {Vecchio}, A.},
        title = "{All correlations must die: Assessing the significance of a stochastic gravitational-wave background in pulsar timing arrays}",
      journal = {\prd},
         year = 2017,
        month = feb,
       volume = {95},
       number = {4},
          eid = {042002},
        pages = {042002},
          doi = {10.1103/PhysRevD.95.042002},
archivePrefix = {arXiv},
       eprint = {1606.09180},
 primaryClass = {astro-ph.IM},
       adsurl = {https://ui.adsabs.harvard.edu/abs/2017PhRvD..95d2002T}
}

@ARTICLE{2016_Arzoumanian,
       author = {{Arzoumanian}, Z. and {Brazier}, A. and {Burke-Spolaor}, S. and {Chamberlin}, S.~J. and {Chatterjee}, S. and {Christy}, B. and {Cordes}, J.~M. and {Cornish}, N.~J. and {Crowter}, K. and {Demorest}, P.~B. and {Deng}, X. and {Dolch}, T. and {Ellis}, J.~A. and {Ferdman}, R.~D. and {Fonseca}, E. and {Garver-Daniels}, N. and {Gonzalez}, M.~E. and {Jenet}, F. and {Jones}, G. and {Jones}, M.~L. and {Kaspi}, V.~M. and {Koop}, M. and {Lam}, M.~T. and {Lazio}, T.~J.~W. and {Levin}, L. and {Lommen}, A.~N. and {Lorimer}, D.~R. and {Luo}, J. and {Lynch}, R.~S. and {Madison}, D.~R. and {McLaughlin}, M.~A. and {McWilliams}, S.~T. and {Mingarelli}, C.~M.~F. and {Nice}, D.~J. and {Palliyaguru}, N. and {Pennucci}, T.~T. and {Ransom}, S.~M. and {Sampson}, L. and {Sanidas}, S.~A. and {Sesana}, A. and {Siemens}, X. and {Simon}, J. and {Stairs}, I.~H. and {Stinebring}, D.~R. and {Stovall}, K. and {Swiggum}, J. and {Taylor}, S.~R. and {Vallisneri}, M. and {van Haasteren}, R. and {Wang}, Y. and {Zhu}, W.~W. and {NANOGrav Collaboration}},
        title = "{The NANOGrav Nine-year Data Set: Limits on the Isotropic Stochastic Gravitational Wave Background}",
      journal = {\apj},
         year = 2016,
        month = apr,
       volume = {821},
       number = {1},
          eid = {13},
        pages = {13},
          doi = {10.3847/0004-637X/821/1/13},
archivePrefix = {arXiv},
       eprint = {1508.03024},
 primaryClass = {astro-ph.GA},
       adsurl = {https://ui.adsabs.harvard.edu/abs/2016ApJ...821...13A}
}

@misc{enterprise,
  author       = {Justin A. Ellis and Michele Vallisneri and Stephen R. Taylor and Paul T. Baker},
  title        = {ENTERPRISE: Enhanced Numerical Toolbox Enabling a Robust PulsaR Inference SuitE},
  month        = sep,
  year         = 2020,
  howpublished = {Zenodo},
  doi          = {10.5281/zenodo.4059815},
  url          = {https://doi.org/10.5281/zenodo.4059815}
}

@misc{enterprise_extensions,
  author       = {Stephen R. Taylor and Paul T. Baker and Jeffrey S. Hazboun and Joseph Simon and Sarah J. Vigeland},
  title        = {ENTERPRISE\_EXTENSIONs},
  year         = {2021},
  url          = {https://github.com/nanograv/enterprise_extensions},
  note         = {v2.4.3}
}

@misc{justin_ellis_2017_1037579,
  author       = {Justin Ellis and
                  Rutger van Haasteren},
  title        = {jellis18/PTMCMCSampler: Official Release},
  month        = oct,
  year         = 2017,
  doi          = {10.5281/zenodo.1037579},
  url          = {https://doi.org/10.5281/zenodo.1037579}
}

@ARTICLE{2021_hui,
       author = {{Hui}, Lam},
        title = "{Wave Dark Matter}",
      journal = {\araa},
         year = 2021,
        month = sep,
       volume = {59},
        pages = {247-289},
          doi = {10.1146/annurev-astro-120920-010024},
archivePrefix = {arXiv},
       eprint = {2101.11735},
 primaryClass = {astro-ph.CO},
       adsurl = {https://ui.adsabs.harvard.edu/abs/2021ARA&A..59..247H}
}

@ARTICLE{1983_Hellings,
       author = {{Hellings}, R.~W. and {Downs}, G.~S.},
        title = "{Upper limits on the isotropic gravitational radiation background from pulsar timing analysis.}",
      journal = {\apjl},
         year = 1983,
        month = feb,
       volume = {265},
        pages = {L39-L42},
          doi = {10.1086/183954},
       adsurl = {https://ui.adsabs.harvard.edu/abs/1983ApJ...265L..39H}
}

@ARTICLE{2008_Burgess,
       author = {{Burgess}, C.~P. and {Conlon}, J.~P. and {Hung}, L. -Y. and {Kom}, C.~H. and {Maharana}, A. and {Quevedo}, F.},
        title = "{Continuous global symmetries and hyperweak interactions in string compactifications}",
      journal = {Journal of High Energy Physics},
         year = 2008,
        month = jul,
       volume = {2008},
       number = {7},
          eid = {073},
        pages = {073},
          doi = {10.1088/1126-6708/2008/07/073},
archivePrefix = {arXiv},
       eprint = {0805.4037},
 primaryClass = {hep-th},
       adsurl = {https://ui.adsabs.harvard.edu/abs/2008...07..073B}
}

@ARTICLE{2009_Goodsell,
       author = {{Goodsell}, Mark and {Jaeckel}, Joerg and {Redondo}, Javier and {Ringwald}, Andreas},
        title = "{Naturally light hidden photons in LARGE volume string compactifications}",
      journal = {Journal of High Energy Physics},
         year = 2009,
        month = nov,
       volume = {2009},
       number = {11},
          eid = {027},
        pages = {027},
          doi = {10.1088/1126-6708/2009/11/027},
archivePrefix = {arXiv},
       eprint = {0909.0515},
 primaryClass = {hep-ph},
       adsurl = {https://ui.adsabs.harvard.edu/abs/2009JHEP...11..027G}
}

@ARTICLE{2011_Cicoli,
       author = {{Cicoli}, Michele and {Goodsell}, Mark and {Jaeckel}, Joerg and {Ringwald}, Andreas},
        title = "{Testing string vacua in the lab: from a hidden CMB to dark forces in flux compactifications}",
      journal = {Journal of High Energy Physics},
         year = 2011,
        month = jul,
       volume = {2011},
          eid = {114},
        pages = {114},
          doi = {10.1007/JHEP07(2011)114},
archivePrefix = {arXiv},
       eprint = {1103.3705},
 primaryClass = {hep-th},
       adsurl = {https://ui.adsabs.harvard.edu/abs/2011JHEP...07..114C}
}

@ARTICLE{2010_Shannon,
       author = {{Shannon}, Ryan M. and {Cordes}, James M.},
        title = "{Assessing the Role of Spin Noise in the Precision Timing of Millisecond Pulsars}",
      journal = {\apj},
         year = 2010,
        month = dec,
       volume = {725},
       number = {2},
        pages = {1607-1619},
          doi = {10.1088/0004-637X/725/2/1607},
archivePrefix = {arXiv},
       eprint = {1010.4794},
 primaryClass = {astro-ph.SR},
       adsurl = {https://ui.adsabs.harvard.edu/abs/2010ApJ...725.1607S}
}

@ARTICLE{Smarra_2024,
       author = {{Smarra}, Clemente and {Kuntz}, Adrien and {Barausse}, Enrico and {Goncharov}, Boris and {Nacir}, Diana L{\'o}pez and {Blas}, Diego and {Shao}, Lijing and {Antoniadis}, J. and {Champion}, D.~J. and {Cognard}, I. and {Guillemot}, L. and {Hu}, H. and {Keith}, M. and {Kramer}, M. and {Liu}, K. and {Perrodin}, D. and {Sanidas}, S.~A. and {Theureau}, G.},
        title = "{Constraints on conformal ultralight dark matter couplings from the European Pulsar Timing Array}",
      journal = {\prd},
         year = 2024,
        month = aug,
       volume = {110},
       number = {4},
          eid = {043033},
        pages = {043033},
          doi = {10.1103/PhysRevD.110.043033},
archivePrefix = {arXiv},
       eprint = {2405.01633},
 primaryClass = {astro-ph.HE},
       adsurl = {https://ui.adsabs.harvard.edu/abs/2024PhRvD.110d3033S}
}

@ARTICLE{pptadr2_timing,
       author = {{Reardon}, D.~J. and {Shannon}, R.~M. and {Cameron}, A.~D. and {Goncharov}, B. and {Hobbs}, G.~B. and {Middleton}, H. and {Shamohammadi}, M. and {Thyagarajan}, N. and {Bailes}, M. and {Bhat}, N.~D.~R. and {Dai}, S. and {Kerr}, M. and {Manchester}, R.~N. and {Russell}, C.~J. and {Spiewak}, R. and {Wang}, J.~B. and {Zhu}, X.~J.},
        title = "{The Parkes pulsar timing array second data release: timing analysis}",
      journal = {\mnras},
         year = 2021,
        month = oct,
       volume = {507},
       number = {2},
        pages = {2137-2153},
          doi = {10.1093/mnras/stab1990},
archivePrefix = {arXiv},
       eprint = {2107.04609},
 primaryClass = {astro-ph.HE},
       adsurl = {https://ui.adsabs.harvard.edu/abs/2021MNRAS.507.2137R}
}

@ARTICLE{eptadr1_timing,
       author = {{Desvignes}, G. and {Caballero}, R.~N. and {Lentati}, L. and {Verbiest}, J.~P.~W. and {Champion}, D.~J. and {Stappers}, B.~W. and {Janssen}, G.~H. and {Lazarus}, P. and {Os{\l}owski}, S. and {Babak}, S. and {Bassa}, C.~G. and {Brem}, P. and {Burgay}, M. and {Cognard}, I. and {Gair}, J.~R. and {Graikou}, E. and {Guillemot}, L. and {Hessels}, J.~W.~T. and {Jessner}, A. and {Jordan}, C. and {Karuppusamy}, R. and {Kramer}, M. and {Lassus}, A. and {Lazaridis}, K. and {Lee}, K.~J. and {Liu}, K. and {Lyne}, A.~G. and {McKee}, J. and {Mingarelli}, C.~M.~F. and {Perrodin}, D. and {Petiteau}, A. and {Possenti}, A. and {Purver}, M.~B. and {Rosado}, P.~A. and {Sanidas}, S. and {Sesana}, A. and {Shaifullah}, G. and {Smits}, R. and {Taylor}, S.~R. and {Theureau}, G. and {Tiburzi}, C. and {van Haasteren}, R. and {Vecchio}, A.},
        title = "{High-precision timing of 42 millisecond pulsars with the European Pulsar Timing Array}",
      journal = {\mnras},
         year = 2016,
        month = may,
       volume = {458},
       number = {3},
        pages = {3341-3380},
          doi = {10.1093/mnras/stw483},
archivePrefix = {arXiv},
       eprint = {1602.08511},
 primaryClass = {astro-ph.HE},
       adsurl = {https://ui.adsabs.harvard.edu/abs/2016MNRAS.458.3341D}
}

@ARTICLE{2023_ding_vlbi,
       author = {{Ding}, H. and {Deller}, A.~T. and {Stappers}, B.~W. and {Lazio}, T.~J.~W. and {Kaplan}, D. and {Chatterjee}, S. and {Brisken}, W. and {Cordes}, J. and {Freire}, P.~C.~C. and {Fonseca}, E. and {Stairs}, I. and {Guillemot}, L. and {Lyne}, A. and {Cognard}, I. and {Reardon}, D.~J. and {Theureau}, G.},
        title = "{The MSPSR{\ensuremath{\pi}} catalogue: VLBA astrometry of 18 millisecond pulsars}",
      journal = {\mnras},
         year = 2023,
        month = mar,
       volume = {519},
       number = {4},
        pages = {4982-5007},
          doi = {10.1093/mnras/stac3725},
archivePrefix = {arXiv},
       eprint = {2212.06351},
 primaryClass = {astro-ph.HE},
       adsurl = {https://ui.adsabs.harvard.edu/abs/2023MNRAS.519.4982D}
}

@misc{2025_Schive,
    author = "Schive, Hsi-Yu",
    title = "{Fuzzy dark matter simulations}",
    eprint = "2509.23231",
    archivePrefix = "arXiv",
    primaryClass = "astro-ph.CO",
    month = "9",
    year = "2025"
}
\bibliographystyle{apsrev4-1.bst}


\newpage

\appendix

\section{Pulsar distances}
\label{pulsar_distance}
Pulsar distance priors $p(L)$ are derived from two types of observations: parallax (PX) measurements and dispersion measure (DM) estimates. We prioritize parallax-based distance measurements from both Very Long Baseline Interferometry (VLBI) and pulsar timing observations. The DM-derived distance serves as a fallback, used only when neither of the two parallax methods provide a reliable estimate.

From the measured PX value a Gaussian prior distribution is derived as,

\beq
p(L) = \frac{1}{\sqrt{2\pi}\sigma_{\varpi}L^{2}} \exp\left[ \frac{- ( \mrm{PX} - L^{-1})^{2}}{2\sigma_{\varpi}^{2}} \right] \ L > 0
\eeq

where PX is the mean distance and $\sigma_{\varpi}$ is its associated uncertainty.

For pulsars with DM-based distance estimates, we adopt a broad, nearly uniform prior with 20\% uncertainty in the center and a half-Gaussian on both sides to account for the asymmetric distribution, the DM-distance prior can be expressed as,

\beq
p(L) = \left\{ 
\begin{aligned}
& \mrm{half \ Gaussian} & 0 < L < 0.8\,L_{\mrm{d}} \\
& \mrm{Uniform} & 0.8\,L_{\mrm{d}} \leq L \leq 1.2\,L_{\mrm{d}}\\
& \mrm{half \ Gaussian} & L > 1.2\,L_{\mrm{d}}
\end{aligned}
\right.
\eeq

where $L_{\mrm{d}}$ is the mean distance derived from the DM measurement. 

\bet[t]
    \centering
    \caption{Pulsar distance estimates, associated uncertainties, and adopted priors (PX or DM based). An asterisk ($*$) identifies pulsars present in both the PPTA-DR3 and EPTA-DR2 data sets.}
    \begin{tabular}{lcccc|lcccc}
    \hline
    $\mrm{Name}$ & $ \mrm{Distance \ (kpc)} $ &  $\mrm{Error (kpc)}$  & $\mrm{Prior}$ & $\mrm{Refs}$ & 
    $\mrm{Name}$ & $ \mrm{Distance \ (kpc)} $ & $\mrm{Error (kpc)}$  & $\mrm{Prior}$ & $\mrm{Refs}$\\
    \hline
    J0030$+$0451$^*$ &   0.329    &      0.005      &      PX       &  \cite{2023_ding_vlbi} & 
    J0125$-$2327     &   1.2       &     0.2            &    PX           &   \cite{MPTA_timing}                           \\
    J0437$-$4715     &    0.156      &   0.001              &      PX       &  \cite{Reardon_2024}            & 
    J0613$-$0200$^*$ &    0.99        &  0.05          &           PX    &  \cite{eptadr2_timing}    \\   
    J0614$-$3329     &    0.67   &      0.25      &         PX      &  \cite{MPTA_timing}             &
    J0711$-$6830     &     0.11  &        0.02    &         DM      &   \cite{MPTA_timing}            \\
    J0900$-$3144$^*$     &   0.38    &    0.08        &     DM         &   \cite{MPTA_timing}            &
    J1017$-$7156     &   1.81    &       0.36     &         DM      & \cite{MPTA_timing}              \\
    J1022$+$1011$^*$ &     0.72     &   0.017    &      PX    &     \cite{2019_Deller}      &
    J1024$-$0719$^*$ &     1.08     &   0.06     &        PX  &       \cite{2023_ding_vlbi}  \\
    J1045$-$4509 &    0.33      &   0.07    &     DM     &    \cite{MPTA_timing}      &
    J1125$-$6014 &    0.86     &     0.22   &    PX     &   \cite{MPTA_timing}       \\
    J1446$-$4701 &    1.4     &   0.7    &     PX      & \cite{MPTA_timing}   &
    J1545$-$4550 &    1.25     &    0.42  &     PX      & \cite{pptadr3_uwl_timing} \\
    J1600$-$3053$^*$ &   1.754      &  0.1    &   PX       & \cite{2025_liu}   &
    J1603$-$7202 &   1.13      &    0.22  &    DM     & \cite{MPTA_timing}     \\
    J1643$-$1224 &    0.91     &  0.08    &      PX   & \cite{2023_ding_vlbi}    & 
    J1713$+$0747$^*$ &     1.2    &  0.03    &     PX    & \cite{2025_liu}    \\
    J1730$-$2304$^*$ &    0.5     &  0.025    &    PX     & \cite{2023_ding_vlbi}     & 
    J1744$-$1134$^*$ &     0.41    &  0.005    &  PX      & \cite{2025_liu}      \\
    J1832$-$0836 &     1.4    &  0.6   &   PX     & \cite{MPTA_timing}        &
    J1857$+$0943$^*$ &    1.11     &  0.08   &    PX    & \cite{eptadr2_timing}        \\
    J1902$-$5105 &    1.65     &  0.33   &    DM    & \cite{MPTA_timing}         &
    J1909$-$3744$^*$ &    1.06     &   0.02   &   PX    & \cite{eptadr2_timing} \\
    J1933$-$6211 &     1.5    &  0.6    &   PX    & \cite{MPTA_timing}        &
    J1939$+$2134 &    2.857    &   0.247    &   PX    &  \cite{2023_ding_vlbi}  \\
    J2124$-$3358$^*$ &    0.47    &   0.02   &   PX       & \cite{eptadr2_timing} & 
    J2129$-$5721 &     6.16    &  1.23   &     DM     &  \cite{MPTA_timing} \\
    J2145$-$0750 &     0.624   &   0.014   &   PX     &  \cite{2019_Deller}  &
    J2241$-$5236 &    1.1   &  0.07    &    PX     & \cite{MPTA_timing} \\
    \hline
    J0751$+$1807 &  1.47    &   0.11      &   PX     & \cite{2025_liu}  & 
    J1012$+$5037 &   0.877  &   0.031      &     PX   & \cite{2023_ding_vlbi}  \\
    J1455$-$3330 &    1.04  &   0.35      &    PX   & \cite{eptadr2_timing}   &
    J1640$+$2224 &   1.37   &   0.113    &     PX    & \cite{2023_ding_vlbi} \\
    J1738$+$0333 &   2.0   &     0.244   &   PX     & \cite{2023_ding_vlbi}  &
    J1751$-$2857 &   0.79  &    0.43    &    PX    & \cite{eptadr2_timing} \\
    J1801$-$1417 &   1.0  &     0.46 &      PX  & \cite{eptadr2_timing} &
    J1804$-$2717 &   0.8  &  0.3   &    PX      & \cite{eptadr2_timing} \\
    J1911$+$1347 &   2.2  &  0.6   &    PX      &  \cite{eptadr2_timing} &
    J1918$-$0642 &   1.408  &  0.14   &    PX     & \cite{2023_ding_vlbi} \\
    J1843$-$1113 &    1.71    &    0.034        &    DM     & \cite{MPTA_timing} & 
    J2322$+$2057 &    0.80     &   0.21        &    PX    & \cite{MPTA_timing} \\
    J1910$+$1256 &   2.778      &    0.48       &    PX    & \cite{2023_ding_vlbi} &
       &      &     &   & \\
    \hline
\end{tabular}
\begin{tablenotes}
	\footnotesize
	\item Note: The DM-derived distances are calculated using the YMW16 model \cite{2017_yao} with an assigned uncertainty of 20\%. The 20\% uncertainty adopted for the DM-based distances represents a simplifying assumption, while the pulsar distances from Earth inherently carry substantial observational uncertainties. Nevertheless, we do not expect the specific choice of distance estimate and prior to significantly impact the derived upper limits.
\end{tablenotes}
\label{pulsar_distance_prior}
\eet

\section{Comparison of EPTA-DR2 results}
\label{eptadr2_compare}

In this appendix, we compare the results of our analysis with previous works from the EPTA. Our updated constraints from the EPTA-DR2 ({\tt{DR2full}}) are shown as the green curve in \Fig{eptadr2_full_uldm_compare}, alongside the limits from \cite{2023_EPTA_uldm} (orange and blue curves). We have also performed a search for scalar ULDM and DPDM using the EPTA-DR2 ({\tt{DR2new}}), following the same methodology and applying the same customized pulsar noise models as in our main analysis. The 95\% upper limits derived for scalar ULDM and DPDM from this EPTA-DR2 ({\tt{DR2new}}) analysis are presented in \Fig{eptadr2_new_uldm} and \Fig{eptadr2_new_dpdm}, respectively.

Our EPTA-DR2 results for scalar ULDM are weaker than those reported in previous studies \cite{2023_EPTA_uldm,2024_epta_4p}. This difference can be attributed to our adoption of a customised pulsar noise model and the use of a linear-uniform prior to derive the upper limits. Regarding the EPTA-DR2 ({\tt{DR2new}}), its shorter observational baseline compared to EPTA-DR2 ({\tt{DR2full}}) results in less stringent constraints, as expected for a dataset covering a reduced time span.

\bef[t]
\centering
\includegraphics[width=0.45\textwidth]{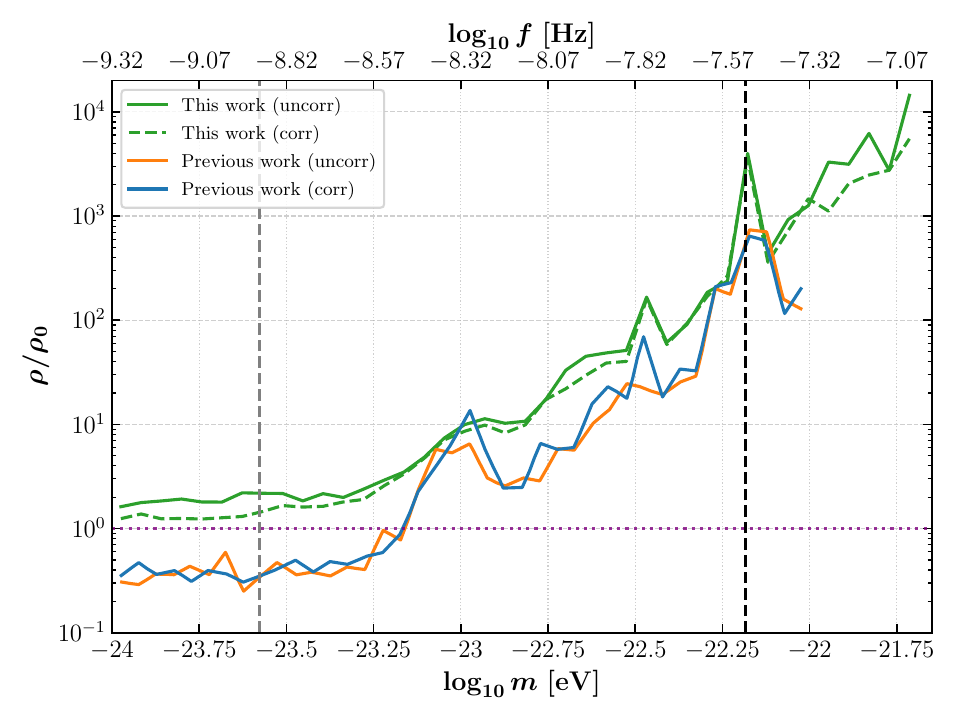}
\includegraphics[width=0.45\textwidth]{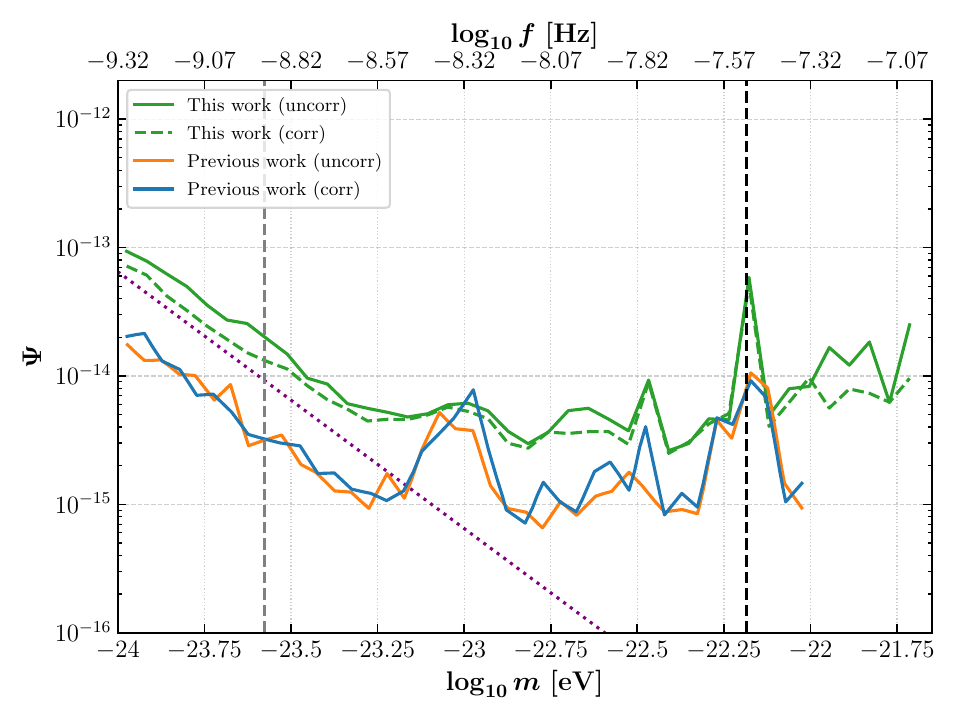}
\caption{Upper limits on scalar ULDM parameters from the analysis of the EPTA-DR2 ({\tt{DR2full}}). The left panel shows constraints on the local dark matter density, while the right panel shows the corresponding limits on the dimensionless oscillation amplitude. The labels “uncorr” and “corr” refer to the uncorrelated and correlated analyses, respectively. The green curves represent the constraints derived in this work from EPTA-DR2 ({\tt{DR2full}}). For comparison, the orange and blue curves show the results from the same dataset reported in \cite{2023_EPTA_uldm}. The purple dotted curve indicates the reference values assuming the local dark matter density of $\rho = 0.4\ \mrm{GeV/cm^3}$. The vertical gray dashed line marks the lowest frequency corresponding to the approximate observational timespan of the EPTA-DR2 ({\tt{DR2full}}) ($\sim$ 24.7 yr), with the black dashed line marking the one-over-one-year reference frequency.}
\label{eptadr2_full_uldm_compare}
\eef

\bef[t]
\centering
\includegraphics[width=0.45\textwidth]{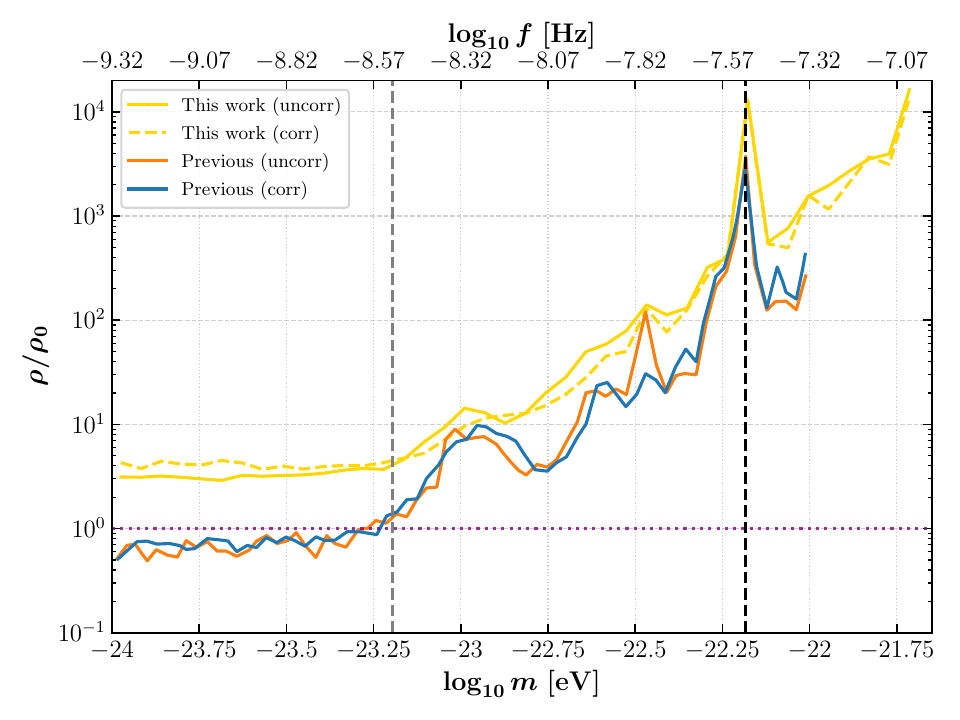}
\includegraphics[width=0.45\textwidth]{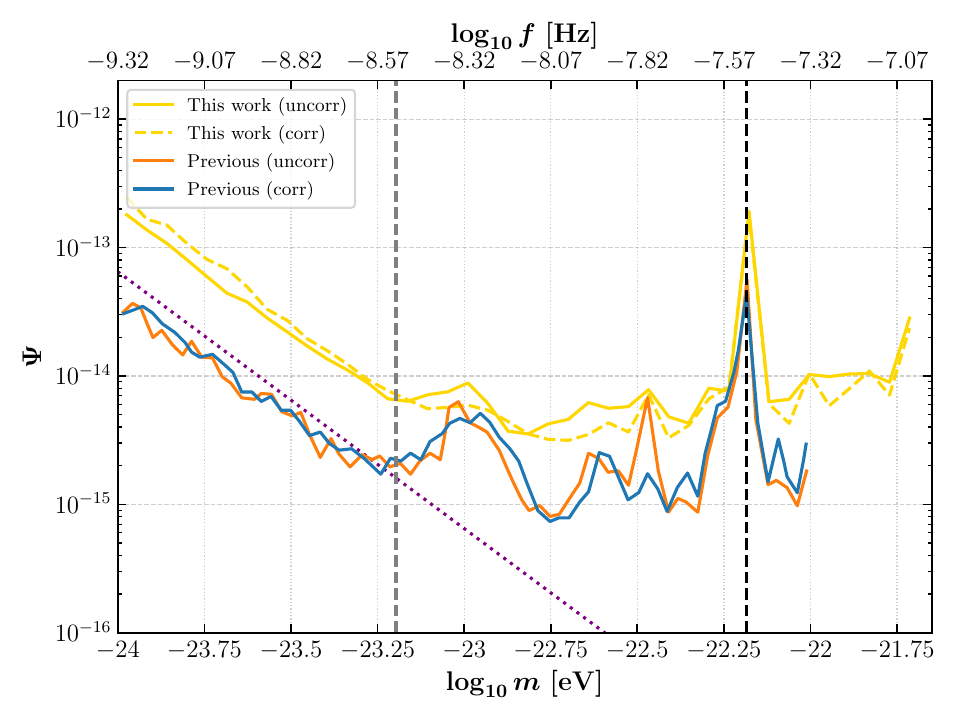}
\caption{Upper limits on scalar ULDM parameters from the analysis of the EPTA-DR2 ({\tt{DR2new}}). The left panel shows constraints on the local dark matter density, while the right panel shows the corresponding limits on the dimensionless oscillation amplitude. The labels “uncorr” and “corr” refer to the uncorrelated and correlated analyses, respectively. The yellow curves represent the constraints derived in this work from EPTA-DR2 ({\tt{DR2new}}). For comparison, the orange and blue curves show the results from the same dataset reported in \cite{2024_epta_4p}. The purple dotted curve indicates the reference values assuming the local dark matter density of $\rho = 0.4\ \mrm{GeV/cm^3}$. The vertical gray dashed line marks the lowest frequency corresponding to the approximate observational timespan of the EPTA-DR2 ({\tt{DR2new}}) ($\sim$ 10.3 yr), with the black dashed line marking the one-over-one-year reference frequency.}
\label{eptadr2_new_uldm}
\eef

\bef[t]
\centering
\includegraphics[width=0.45\textwidth]{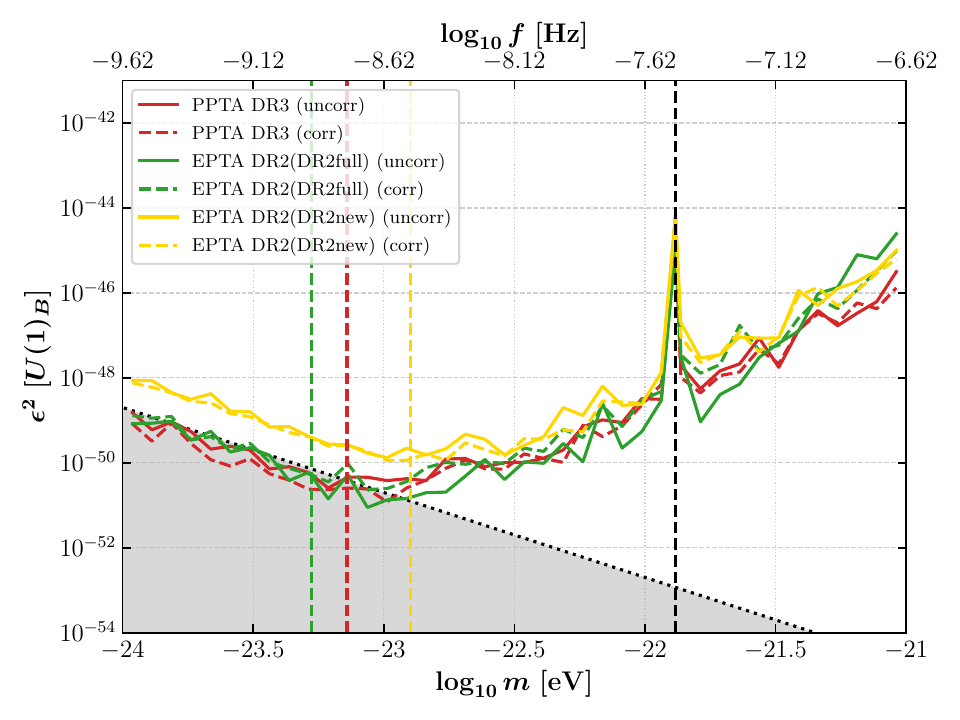}
\includegraphics[width=0.45\textwidth]{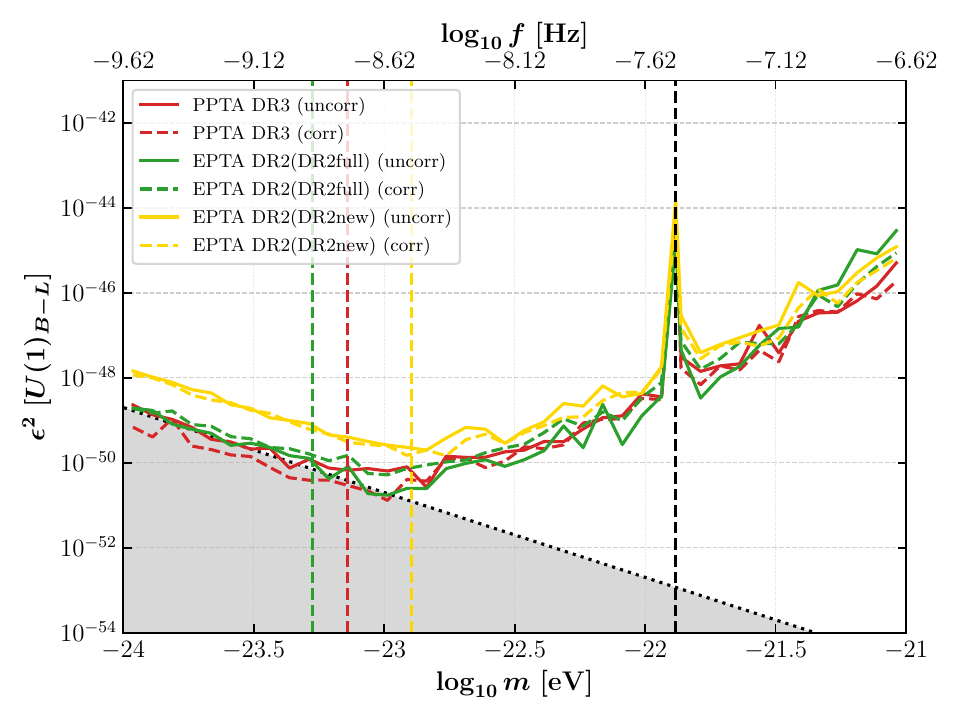}
\caption{Constraints on the dark photon coupling strength $\epsilon^2$ for the $U(1)_B$ (left) and $U(1)_{B-L}$ (right) interactions from this analysis, including results from PPTA-DR3 (red), EPTA-DR2 ({\tt{DR2full}}) (green) and EPTA-DR2 ({\tt{DR2new}}) (yellow). Solid and dashed curves represent the uncorrelated and correlated cases, respectively. The black dotted line and gray shaded region indicate the parameter space where gravitational effects dominate over fifth-force couplings, based on the simplified assumption that vector ULDM produces gravitational oscillations three times stronger than scalar ULDM. Vertical dashed lines indicate the approximate observational timespans: PPTA-DR3 ($\sim$18 yr), EPTA-DR2 ({\tt{DR2full}}) ($\sim$24.7 yr) and EPTA-DR2 ({\tt{DR2new}}) ($\sim$ 10.3 yr), with the black dashed line marking the one-over-one-year reference frequency.}
\label{eptadr2_new_dpdm}
\eef

\end{document}